\documentclass[12pt]{article}
\usepackage{ccaption}
\usepackage{amssymb}
\usepackage{gensymb}
\usepackage{times}
\usepackage{titling}
\usepackage{graphicx}
\usepackage{amsmath}
\usepackage{multibib}
\usepackage[version=3]{mhchem}
\usepackage{color}
\usepackage{bm}
\usepackage{lineno}
\usepackage{multirow}
\usepackage{xr}
\usepackage{float}

\usepackage[paper=letterpaper,twoside,top=2.0cm,bottom=2cm,left=1.5cm,
            right=1.5cm,bindingoffset=0.0cm,voffset=0.0cm]{geometry}
\usepackage[labelfont=bf,labelsep=period]{caption}
\usepackage[symbol]{footmisc}
\usepackage[pagebackref=false]{hyperref}
\hypersetup{
    colorlinks=true,
    linkcolor=blue,
    citecolor={blue}}
    
\renewcommand{\refname}{\subsection*{References}}

\renewcommand{\thesection}{}
\renewcommand{\thesubsection}{\arabic{subsection}}  

\title{
{\bf Chemical Design of Electronic and Magnetic Energy Scales in Tetravalent Praseodymium} \\
}
\author{
 	Arun Ramanathan$^{1}$,\, Jensen Kaplan$^{2}$,\, Dumitru-Claudiu
     Sergentu$^{3,4,9}$,\, Jacob A. Branson$^{5,8}$\\ Mykhaylo Ozerov$^{6}$,\, Alexander I. Kolesnikov$^{7}$,\,Stefan G. Minasian$^{8}$,\, Jochen Autschbach$^{9}$,\\ John W. Freeland$^{10}$,\,Zhigang Jiang$^2$,\, Martin Mourigal$^{2}$, Henry S. La Pierre$^{1,11,}$\thanks{Email: hsl@chemistry.gatech.edu}\\
	\normalsize{$^1$School of Chemistry and Biochemistry, Georgia Institute of Technology, Atlanta, GA 30332, USA}
    \\
    \normalsize{$^2$School of Physics, Georgia Institute of Technology, Atlanta, GA 30332, USA} 
	\\
        \normalsize{$^3$University of Rennes, CNRS, Institut des Sciences Chimiques de Rennes - UMR 6226, F-35000 Rennes, France}
        \\
        \normalsize{$^4$A. I. Cuza University of Iași, RA-03 Laboratory (RECENT AIR), Iași 700506, Romania}
        \\
 	\normalsize{$^5$Department of Chemistry, University of California, Berkeley, California 94720, United States}
	\\
	\normalsize{$^6$National High Magnetic Field Laboratory, Florida State University, Tallahassee, Florida 32310, USA}
	\\
	\normalsize{$^7$Neutron Scattering Division, Oak Ridge National Laboratory, Oak Ridge, TN 37831, USA} 
	\\
        \normalsize{$^8$Lawrence Berkeley National Laboratory, Berkeley, California 94720, USA}
        \\
	\normalsize{$^9$Department of Chemistry, University at Buffalo, State University of New York, Buffalo, NY 14260-3000, USA}
	\\
    \normalsize{$^{10}$Advanced Photon Source, Argonne National Laboratory, Lemont, IL, 60439, USA}
        \\
	\normalsize{$^{11}$Nuclear and Radiological Engineering and Medical Physics Program, School of Mechanical Engineering},\\
    \normalsize{Georgia Institute of Technology, Atlanta, Georgia 30332, USA}
    \\
	} 
\date{\today}
\newcites{Methods}{References}

\begin{document}
\maketitle


\section*{Abstract} \baselineskip20pt 
\baselineskip20pt

{\bf Lanthanides in the trivalent oxidation state are typically described using an ionic picture that leads to localized magnetic moments. The hierarchical energy scales associated with trivalent lanthanides produce desirable properties for e.g., molecular magnetism, quantum materials, and quantum transduction. Here, we show that this traditional ionic paradigm breaks down for praseodymium in the 4+ oxidation state. Synthetic, spectroscopic, and theoretical tools deployed on several solid-state Pr$^{4+}$-oxides uncover the unusual participation of $4f$ orbitals in bonding and the anomalous hybridization of the $4f^1$ configuration with ligand valence electrons, analogous to transition metals. The resulting competition between crystal-field and spin-orbit-coupling interactions fundamentally transforms the spin-orbital magnetism of Pr$^{4+}$, which departs from the $J_{\rm eff}\!=\!1/2$ limit and resembles that of high-valent actinides. Our results show that Pr$^{4+}$ ions are in a class on their own, where the hierarchy of single-ion energy scales can be tailored to explore new correlated phenomena in quantum materials.}

\section*{Introduction} 

The electronic structure of lanthanide and actinide materials inherits on-site correlations and unquenched orbital degrees of freedom from atomic $f$-electron states. In the most stable trivalent oxidation state (Ln$^{3+}$), the core-like 4$f$ orbitals are only weakly perturbed by the crystalline environment \cite{newman1971theory}. Yet, the energetic splitting of the otherwise 2$J$+1 fold-degenerate (free-ion) ground-state yields rich physics and applications. For example, Ln$^{3+}$ insulators can host anisotropic magnetic moments with effective spin-1/2 character ($J_{\rm eff}\!=\!1/2$) that are promising to stabilize entangled states in quantum magnets \cite{kimura2013quantum, gao2019experimental}. Metallic 4$f$ and 5$f$ systems also display a wealth of quantum phenomena rooted in the hybridization between localized $f$-electrons and conduction $d$-electrons such as the Kondo effect, valence fluctuations, correlated insulators, and unconventional superconductivity \cite{coleman1984new, lawrence1981valence}. 

In rare instances, Ce, Pr, Tb, (and less definitively Nd, and Dy) ions exist in a high, formally tetravalent, oxidation state, \textit{i.e.} Ce$^{4+}$ (4$f^0$), Pr$^{4+}$ (4$f^1$), and Tb$^{4+}$ (4$f^7$) \cite{gompa2020chemical,meyer1990synthesis}. Although covalent metal-ligand interactions involving the 4f shell are generally weak in Ln$^{3+}$ systems \cite{crosswhite1980spectrum,cotton2013lanthanide}, this paradigm breaks down for Ln$^{4+}$ as 4$f$ orbitals directly participate in bonding and anomalously hybridize with the valence orbitals of the ligands (\textit{e.g.}\ the 2$p$ states for oxygen) analogous to transition metals. A high oxidation state and strong 4f covalency are expected to significantly impact the redox, electronic, and magnetic properties of these systems, but, surprisingly, only a few Ln$^{4+}$ insulators have been studied in detail to date \cite{kern1985crystal,boothroydPr02}. As Ce$^{4+}$ is nominally non-magnetic and Tb$^{4+}$ has a half-filled 4$f$ shell, the one-electron $4f^1$ configuration of Pr$^{4+}$ makes it unique to search for new quantum phenomena at the nexus of strong electronic correlations, quantum magnetism, and spin-orbital entanglement. 

The emergence of an insulating state in PrBa$_2$Cu$_3$O$_{6+\delta}$ (PBCO) -- a compound obtained by substituting Y by Pr in the high-$T_c$ superconductor YBa$_2$Cu$_3$O$_{6+\delta}$ (YBCO) -- epitomizes such anomalous behavior. In PBCO, PrO$_8$ cuboids share oxygen atoms with the CuO$_2$ planes. The significant Pr-4$f$/O-2$p$ covalency (Fehrenbacher-Rice hybridization) drives a mixed-valent state for Pr ions that competes with Cu-3$d$/O-2$p$ hybridization (Zhang-Rice) and dramatically suppresses superconductivity in favor of local magnetism  \cite{fehrenbacher1993unusual}. This observation stimulated early experimental and theoretical work to understand the magnetism of cubic Pr$^{4+}$ oxides such as PrO$_2$ and BaPrO$_3$ \cite{kern1985crystal,boothroydPr02}. More recently, the edge-sharing PrO$_6$ octahedral in Na$_2$PrO$_3$ have attracted attention to stabilize antiferromagnetic Kitaev interactions between $J_{\rm eff}\!=\!1/2$ moments \cite{jang2019antiferromagnetic}. But much like in PrO$_2$~\cite{boothroydPr02}, the hallmark of Na$_2$PrO$_3$ magnetism is the unusually large crystal field (CF) splitting $\Delta_{\rm CF} \geq 230$~meV that competes with spin-orbit coupling (SOC) $\zeta_{\rm SOC} \approx 100$~meV \cite{daum2021collective}. The competition between CF and SOC yields drastically different single-ion and exchange properties than expected in the $\Delta_{\rm CF} \ll \zeta_{\rm SOC}$ limit where $J_{\rm eff}\!=\!1/2$ moments usually form, as illustrated in Fig.~\ref{fig1}a. The most noticeable consequences for Na$_2$PrO$_3$ are the low effective magnetic moment with $g \approx 1$ and the surprisingly large $J \approx 1$ meV exchange interactions \cite{daum2021collective}. The precise mechanisms that endow Pr$^{4+}$ ions with these unusual properties are poorly understood. 

In this work, we focus on the microscopic mechanisms that underpin the electronic and magnetic behavior of Pr$^{4+}$ materials comprising octahedral [PrO$_6$]$^{8-}$ units. We examine a series of insulating oxides with decreasing order of lattice dimensionality: quasi-2D layers in Na$_2$PrO$_3$ (\textbf{2-Pr}, Fig.~\ref{fig1}b, Ref.~\cite{ramanathan2021plane}) and quasi-1D chains in Sr$_2$PrO$_4$ (\textbf{1-Pr}, Fig.~\ref{fig1}c, Ref.~\cite{fiscus2000synthesis}) to understand magnetic exchange and the role of Pr-4$f$/O-2$p$ hybridization; quasi-isolated ``0D'' PrO$_6$ octahedral in Li$_8$PrO$_6$ (\textbf{0-Pr}, Fig.~\ref{fig1}d, Ref.~\cite{wolf1985li8pro6}) to understand the intrinsic behavior of the [PrO$_6$]$^{8-}$ moiety without the complication of magnetic exchange interactions. Using inelastic neutron scattering (INS) and infrared magneto-spectroscopy (IRMS), we demonstrate that the magnetic ground-state of Pr$^{4+}$ ions {\it systematically} deviates from the $J_{\rm eff}\!=\!1/2$ limit and can be understood from an intermediate coupling regime where significant admixture of nominally excited $J$-states enter the ground-state wave-function. X-ray magnetic circular dichroism (XMCD) at the Pr-$M_{4,5}$ edge strengthens that picture and elucidates the mechanism behind the low effective magnetic moments of Pr$^{4+}$ ions. Oxygen $K$-edge and Pr $M_{4,5}$-edge x-ray absorption spectroscopy (XAS) evidences Pr-4$f$/O-2$p$ hybridization with a degree of Pr-O covalency influenced by the symmetry of the [PrO$_6$]$^{8-}$ moiety. These results are supported and explained by first-principles calculations. Taken together, this study establishes Pr$^{4+}$ ions as an important building block to design quantum magnetic behavior distinct from trivalent lanthanides. Furthermore, the study demonstrates that the competition between energy scales in the [PrO$_6$]$^{8-}$ unit is reminiscent of 4$d$ and 5$d$ transition metals~\cite{bahrami2022first}, and can serve as a simplification of 5$f^1$ actinide systems for which $\Delta_{\rm CF}$,  $\zeta_{\rm SOC}$ and on-site Hubbard interaction $U$ strongly compete~\cite{zhou2012self,santini2009multipolar}.

\section*{Results}
Crystalline powder samples of \textbf{0-Pr}, \textbf{1-Pr}, and \textbf{2-Pr} were synthesized using solid-state reactions and phase purity was confirmed using powder X-ray diffraction (See Methods and Supplementary Sections 1 and 2). These materials incorporate low symmetry, but close-to-octahedral [PrO$_6$]$^{8-}$ units with $D_{2d}$ symmetry in \textbf{2-Pr} ($C_{2/c}$ space group, Fig.~\ref{fig1}b), $C_{2h}$ in \textbf{1-Pr} ($Pbam$ space group, Fig.~\ref{fig1}c), and $S_6$ in \textbf{0-Pr} ($R\overline{3}$ space group, Fig.~\ref{fig1}d). \textbf{0-Pr} contains spatially isolated PrO$_6$ octahedral with a nearest neighbor Pr-Pr distance of $d\approx5.6$~\AA, which is significantly longer than $d\approx3.5$~\AA\ in \textbf{1-Pr} and \textbf{2-Pr}, and sufficient to effectively magnetically isolate the [PrO$_6$]$^{8-}$ units.

Broad-band INS measurements were used to probe the CF states accessible to the dipole selection rule. Given that Pr$^{4+}$ is a 4$f^1$ Kramers ion (isoelectronic to Ce$^{3+}$), the standard approach starts from a $^2F$ free-ion manifold split by SOC into $J\!=\!5/2$ and $J\!=\!7/2$ multiplets. For a six-oxygen environment with $O_h$ symmetry, the CF further splits the $^2F_{5/2}$ multiplet into a doublet ground-state ($\Gamma_7$) and an excited quartet ($\Gamma_8$) and the $^2F_{7/2}$ mutiplet into two doublets ($\Gamma_7^\prime$ and $\Gamma_6$), and a quartet ($\Gamma_8^\prime$). Any deviation from $O_h$ symmetry, as is the case for our materials (see Fig~\ref{fig1}), splits the quartets and yields seven Kramers doublets (KDs). Thus, the magnetic properties of Pr$^{4+}$ ions in the hypothetical $\Delta_{\rm CF} \ll \zeta_{\rm SOC}$ limit are dominated by the $\Gamma_7$ doublet ground-state, which is spanned by pseudospin variable $|\pm\rangle$ associated with an effective angular momentum $J_{\rm eff}\!=\!1/2$. The wave-function of the $\Gamma_7$ doublet is well-known~\cite{jang2019antiferromagnetic} and can be written in either $|J,m_J\rangle$ or $|m_l,m_s\rangle$ basis (see Fig~\ref{fig1}a and SI). 

However, as $\zeta_{\rm SOC} \approx \Delta_{\rm CF}$ for Pr$^{4+}$, the simple $J_{\rm eff}\!=\!1/2$ picture breaks down. Indeed, INS on \textbf{0-Pr} readily reveals an intense magnetic signal at $E_1^{\rm \textbf{0-Pr}}=274(1)$~meV which we assign to the \textit{first} CF excitation, see Fig.~\ref{fig2}a. This energy is 2.5 times larger than reported for isoelectronic Ce$^{3+}$ in KCeO$_2$~\cite{bordelon2021magnetic,eldeeb2020energy}, and to the best of our knowledge, this is the largest first CF excited state observed for a lanthanide ion. Given the uncommonly large $\Delta_{\rm CF}$, modeling the single-ion properties of \textbf{0-Pr} requires an intermediate coupling approach~\cite{boothroydPr02} that uses the complete set of 14 $|m_l,m_s\rangle$ basis states and diagonalizes the single-ion CF Hamiltonian $\mathcal{\hat{H}}_{\rm CF}^{\rm Pr}=B^0_2 \hat{O}^0_2 + B^0_4\hat{O}^0_4+ B^4_4\hat{O}^4_4 + B^0_6\hat{O}^0_6 + B^4_6\hat{O}^4_6$ for a fixed value of the spin-orbit interaction $\zeta_{\rm SOC}$ (see Methods and Suppl. Sec. 3 for definitions). The above single-ion CF Hamiltonian is written in a \emph{truncated} symmetry basis to avoid over-parametrization and treat all materials on equal footing (see Methods). Irrespective, it is impossible to constrain the parameters of $\mathcal{\hat{H}}_{\rm CF}^{\rm \textbf{0-Pr}}$ solely using the one observed excitation. Thus \textbf{1-Pr} is examined first because the PrO$_6$ octahedra further depart from ideal symmetry and are likely to present a richer spectrum in INS.

Unlike \textbf{0-Pr}, INS results on \textbf{1-Pr} reveal three magnetic excitations at $E_1^{\rm \textbf{1-Pr}}=168(1)$ meV, $E_2^{\rm \textbf{1-Pr}}=335(1)$ meV, and $E_3^{\rm \textbf{1-Pr}}=387(1)$ meV (Fig.~\ref{fig2}d); more states than available in the sole $J\!=\!5/2$ manifold. Although \textbf{1-Pr} exhibits an antiferromagnetic transition at $T_{\rm N} = 3.0$~K with a pronounced peak in $\chi(T)$ (Fig.~\ref{fig2}e), the magnetic susceptibility at $\mu_0H = 3$ T above $T>40$~K can be used to further constrain the parameters of the single-ion CF Hamiltonian. To proceed, $\mathcal{\hat{H}}_{\rm CF}^{\rm Pr}$ is diagonalized with fixed $\zeta_{\rm SOC}=112$~meV (free-ion value) and the CF parameters fit to the observed INS energies and magnetic susceptibility data (see Methods and Supplementary Section. 3). This yields a model Hamiltonian that reproduces the isothermal magnetization at $T\!=\!50$ K (Fig.~\ref{fig2}f) and predicts an unusually small powder-averaged $g$ factor $g_{\rm CF}^{\rm avg,\textbf{1-Pr}}=1.13$ and an effective moment $\mu_{\rm CF}^{\rm eff,\textbf{1-Pr}}\!\approx\!1~\mu_{\rm B}$/Pr comparable to the value extracted from a Curie-Weiss fit $\mu_{\rm CW}^{\rm eff,\textbf{1-Pr}}\!=\!1.13(1)~\mu_{\rm B}$/Pr (Fig.~\ref{fig2}e).

Having established an approach to model the single-ion Hamiltonian for Pr$^{4+}$, \textbf{0-Pr} is examined and employ IRMS measurements conducted up to 17.5~T. The normalized IR spectra reveal three field-dependent features around $E_1^{\rm \textbf{0-Pr}}=267$~meV, $E_2^{\rm \textbf{0-Pr}}\!=\!270$~meV, and $E_4^{\rm \textbf{0-Pr}}\!=\!670$~meV (Fig.~\ref{fig2}c) that can be associated with magnetic-dipole allowed CF transitions from the ground-state doublet. The distinct features at $E_1^{\rm \textbf{0-Pr}}$ and $E_2^{\rm \textbf{0-Pr}}$ -- resolved due to the excellent energy resolution of IRMS -- correspond to the sole transition observed in INS. The first excited level in \textbf{0-Pr} is thus a quasi-degenerate quartet ($\Gamma_8$-like) split by the weak distortion of the PrO$_6$ octahedra from an ideal $O_h$ symmetry. This model is fully supported by ab-initio calculations (multireference CASPT2+SOC, Methods and Supplementary Section. 4), which predicts the quasi-degeneracy of $E_2^{\rm \textbf{0-Pr}}$ and $E_2^{\rm \textbf{0-Pr}}$ at $241$~meV and $246$~meV, respectively. The $670$~meV transition is likely weak in INS and masked by the strong background (recoil intensity observed from hydroxide impurities, $<5\%$ in starting materials, see Suppl. Sec. 1). The wavefunction calculations assign it as the \emph{fourth} ground to excited state transition, and further predict a third (IR inactive) transition at $396$ meV with $^2T_{2u} + ^2A_{2u}$ origin~\cite{gelessus1995multipoles}. The parameters of $\mathcal{\hat{H}}_{\rm CF}^{\rm \textbf{1-Pr}}$ are fitted using the same procedure as for \textbf{0-Pr} (Fig.~\ref{fig2}b). The resulting model yields $g_{\rm CF}^{\rm av,\textbf{0-Pr}}$ = 0.94, in good agreement with the isothermal magnetization (Fig.~\ref{fig2}f), and $\mu_{\rm CF}^{\rm eff,\textbf{0-Pr}}\!=\!0.81\mu_{\rm B}$/Pr consistent with $\mu_{\rm CW}^{\rm eff,\textbf{0-Pr}}\!=\!0.86(1)~\mu_{\rm B}$/Pr (Fig.~\ref{fig2}b) and first-principles calculations (See Discussion). Analysis of the INS spectrum of \textbf{2-Pr} leads to similar conclusions (See Ref.~\cite{daum2021collective}).

Analysis of the single-ion physics of these three materials therefore suggests that the ground-state wave-function of Pr$^{4+}$ systematically deviates from the $\Gamma_7$ doublet expected in the $J_{\rm eff}=1/2$ limit. For example, the ratio of $|m_l\!=\!\mp3,m_s\!=\!\pm1/2\rangle$ to $|m_l\!=\!\pm2,m_s\!=\!\pm1/2\rangle$  basis states (parametrized by $(A^2/B^2)$, see Fig.~\ref{fig1} caption and Methods for Definition and Supplementary Tables S5-S7 for full wavefunction) changes from 2.6 for the $|\Gamma_7\rangle$ doublet to 0.53, 2.13, and 1.51 for the ground-state doublet of \textbf{0-Pr}, \textbf{1-Pr}, and \textbf{2-Pr}, respectively. When recast in the $|J,m_J\rangle$ basis, our analysis suggests that intermediate coupling mixes $|J\!=\!7/2,m_J\!=\!\pm3/2\rangle$ and $|J\!=\!7/2,m_J\!=\!\pm5/2\rangle$ states into the ground doublet, leading to an increased $|J,m_J\!=\!\pm5/2\rangle$ character for the ground-state wavefunction (see Suppl. Sec. 3). 

Probing the density of electronic states around the 4$f$ level, X-ray absorption spectroscopy (XAS, see Methods) provides definitive spectroscopic evidence of this hypothesis and further elucidates the origin of the large $\Delta_{\rm CF}$. The $M_5$ (3$d_{5/2}\!\rightarrow\! 4f_{7/2} \ \&\ 4f_{5/2}$) and $M_4$ (3$d_{3/2}\rightarrow$ 4$f_{5/2}$) edges for \textbf{2-Pr} and \textbf{0-Pr} are shown in Fig.~\ref{fig3}(a,b). The splitting between the Pr $M_5$ edge at 931~eV and $M_4$ edge at 951~eV originates from SOC within the 3$d$ core-holes. The $M_{5,4}$ edges for both \textbf{0-Pr} and \textbf{2-Pr} show intense main peaks (labelled $A$ and $B$) followed by higher-energy satellites ($A^{\prime}$ and $B^{\prime}$) raised by $\approx 5$~eV, and a smeared shoulder (labelled $B^S$) starting $\approx 3$~eV below the main peaks. For a 4$f^1$ ion with a ground-state wave-function comprised of a pure $J$-multiplet, isotropic $M_{5,4}$ edges as predicted by the Wigner-Eckart theorem~\cite{finazzi1995direct}(Fig. S4) are expected. Thus, the complex spectral lineshapes observed in \textbf{0-Pr} and \textbf{2-Pr}, which resemble previous observations for PrO$_2$~\cite{minasian2017quantitative} (see Suppl. Fig. S4 and S5), are direct evidence for the mixing of $^2F_{7/2}$ and $^2F_{5/2}$ multiplets in the ground and excited states of our compounds. This analysis is corroborated by the branching ratio (BR) $I_{M_5}/I_{M_5} + I_{M_4}$, evaluated from the total spectral weight under all $M_5$ or $M_4$ peaks, with values of $0.445(17)$ (\textbf{2-Pr}), $0.443(10)$ (\textbf{0-Pr}), and $0.453(6)$ (PrO$_2$). These values are less than $\approx0.50$ reported for ionic Ce$^{3+}$ systems\cite{minasian2017quantitative,finazzi1995direct} (a BR of $\approx0.5$ also applies for an ionic Pr$^{4+}$ system with no hybridization as shown in Fig. S4). 

The complex $M_{5,4}$ lineshapes and multiplet mixing in 4$f^1$ systems have previously been ascribed to electronic hybridization, \textit{i.e.}\ strong covalent bonding, between 4$f$ and O-2$p$ states~\cite{finazzi1995direct,okane2012magnetic}. In the charge-transfer limit (Zaanen-Sawatzky-Allen (ZSA) scheme~\cite{zaanen1985band}), the electronic ground-state of Pr$^{4+}$ is a superposition  $|\psi_g\rangle=\sin\theta~|4f^1\rangle+\cos\theta~|4f^2\underline\nu\rangle$ where $\underline\nu$ is a hole in the O 2$p$ band. Configuration-interaction (CI) calculations using the Anderson-impurity model (AIM)~\cite{anderson1961localized,cowan1981theory} were carried out to understand the impact of 4f covalency on the $M_{5,4}$ XAS spectra of the subject compounds. Within this framework, the initial and final states include the combination of $|3d^{10}4f^1\rangle$ \& $|3d^{10}4f^2\underline\nu\rangle$, and $|3d^94f^2\rangle$ \& $|3d^94f^3\underline\nu\rangle$, respectively (see Methods). In the limit of vanishing hybridization ( $V\!\rightarrow\!0$), the energy difference between initial configurations is $\Delta E_{g}\!=\!2.0$~eV (\textbf{2-Pr}) and $3.0$ eV (\textbf{0-Pr}), and between final states is $\Delta E_{f}=0.5$ eV (\textbf{2-Pr}) and $1.5$ eV (\textbf{0-Pr}). Calculations for a realistic hybridization agree well with the experimental data (Fig.~\ref{fig3}(a,b) and allows to estimate the fraction of $4f^1$ and $4f^2$ configurations in $|\psi_g\rangle$ to be $69\%$-$31\%$ for \textbf{2-Pr}, and $75\%$-$25\%$ in \textbf{0-Pr}; i.e., \textbf{0-Pr} is the least hybridized system. The estimated contributions to $|\psi_g\rangle$ in \textbf{2-Pr} are similar to the values extracted for PrO$_2$ (see Fig. S5)~\cite{minasian2017quantitative}. The smaller hybridization in \textbf{0-Pr} relative to both \textbf{2-Pr} and PrO$_2$ is evident theoretically from the increased $\Delta E_g$ and experimentally from the more dominant main peak at both edges. Due to the comparable energy scales of $\Delta E_g \approx V_g$, Pr$^{4+}$ oxide systems are thus strongly correlated insulators in the charge-transfer limit ($U_{ff} \gg \Delta E_g$) and require a quantum many-body description. Indeed, these systems behave similar to CeO$_2$ and the spectral features are describing ground and excited state charge transfer~\cite{sergentu2021probing}.

The weak magnetic moment observed for the Pr$^{4+}$ ion can also be understood directly from XAS. The XMCD spectra at the Pr $M_{5,4}$ edge (See Methods) which reveal sizeable dichroism (Fig~\ref{fig3}a (\textbf{2-Pr}) and \ref{fig3}b (\textbf{0-Pr})). Quantitative analysis using sum rules (see Suppl. Sec. S3.7) allows for the extraction of the orbital $\mu_{\rm o}\!=\!-\langle{L_z}\rangle ~\mu_{\rm B}$, spin ($\mu_{\rm s}\!=\!-2\langle{S_z}\rangle ~\mu_{\rm B}$, and magnetic dipole $\langle{T_z}\rangle$ contributions to the total moment $\mu_{\rm t}$)~\cite{thole1992x,thole1992x} (see Methods, note that  $\langle{T_z}\rangle\!\neq\!0$  reflects the departure from spherical symmetry for $\mu_{\rm s}$) ). Applying the orbital sum-rule yields 
$\mu_{\rm o}=0.34(5)~\mu_{\rm B}$ (\textbf{2-Pr}) and $0.33(7)~\mu_{\rm B}$ (\textbf{0-Pr}). The measured absolute macroscopic magnetization yields $\mu_{\rm t}\!=\!0.028(4)~\mu_{\rm B}$ (\textbf{2-Pr}) and $0.044(3)~\mu_{\rm B}$ (\textbf{0-Pr}) at $\mu_0H=5$ T and $T=20$ K, and in turn yields $\mu_{\rm s}\!=\!-0.31(9)~\mu_{\rm B}$ (\textbf{2-Pr}) and $-0.29(10)~\mu_{\rm B}$ (\textbf{0-Pr}) based on $\mu_{total}=\mu_{spin}+\mu_{orbital}$. These values correspond to $|\langle{T_z}\rangle/\langle{S_z}\rangle|=0.40(8)$ (\textbf{2-Pr}) and $0.38(1)$ (\textbf{0-Pr}) and
$|\langle{L_z}\rangle/\langle{S_z}\rangle|=2.17(6)$ (\textbf{2-Pr}) and $2.29(9)$ (\textbf{0-Pr}); the latter is significantly lower than $|\langle{L_z^{4f^1}}\rangle/\langle{S_z^{4f^1}}\rangle|=8$ expected for a free $4f^1$ ion~\cite{tripathi2018xmcd} but resembles $5f$--$c$ hybridized uranium systems and is usually attributed to $4f$ electron delocalization in lanthanides~\cite{finazzi1997x,pedersen2019uf6}. Despite a low bulk magnetic moment, XMCD data reveals the existence of sizable spin and orbitals moments with a reduced $|\langle{L_z}\rangle/\langle{S_z}\rangle|$ ratio that provides a fingerprint for Pr-$4f$/O-$2p$ hybridization in Pr$^{4+}$ systems. 

Finally, to gain ligand-based information about Pr-$4f$/O-$2p$ hybridization, O $K$-edge XAS~\cite{frati2020oxygen} were acquired. The spectra for \textbf{2-Pr} and \textbf{0-Pr}, Fig~\ref{fig3}(c,d), reveal strong features in the 532.8~eV to 536~eV range attributed to excitations from the $1s$ shells of the ligand to states with Pr-$5d$ and O-$2p$ character. This is a measure of the $5d$-covalency of the Pr--O bond and shows that nominally unoccupied $5d$ orbitals take part in covalent bonding~\cite{altman2016evidence}. The splitting of the $5d$ states is estimated to be $3.67(11)$ eV in \textbf{2-Pr} and $3.61(4)$ ev in \textbf{0-Pr} and compares well with the value calculated for PrO$_2$ ($\approx3.6$ eV, see Fig. S3)~\cite{minasian2017quantitative}. Contributions from Pr-$6sp$ states cannot be entirely neglected in the $5d$ driven region~\cite{hu19992}. More subtle features common to both \textbf{2-Pr} and \textbf{0-Pr} include pre-$5d$-edge peaks at near $\approx529$ eV and $\approx530.7$~eV (labeled as $1s \rightarrow 4f$). These pre-edge peaks are a signature of strong Pr-$4f$/O-$2p$ hybridization in the ground state ($|\psi_g\rangle$) because they reflect transitions from the O $1s$-core states to $2p$-hole states of the oxygen in the narrow $4f$-dominated bands. These low-energy pre-edge features are characteristic of Ln$^{4+}$ ions; if at all present in spectra of Ln$^{3+}$ systems\cite{altman2016evidence} , they are quite weak. The integrated intensities of the $1s \rightarrow 4f$ peaks is 3.4(1) and 2.5(1) larger for PrO$_2$ and \textbf{2-Pr}, respectively, than for \textbf{0-Pr}. This result indicates that \textbf{0-Pr} has the least Pr-$4f$/O-$2p$ hybridization in good accord with the Pr $M_{5,4}$ edge spectra. Overall, the presence of pre-edge features in the O $K$-edge XAS spectra confirms $4f$ covalency in the Pr-O bond and strongly indicates ligand holes induced by Pr-$4f$/O-$2p$ hybridization~\cite{hu19992}.

\section*{Discussion}
\begin{figure}[h!]
    \begin{center}
        \includegraphics[width=0.75\textwidth]{ 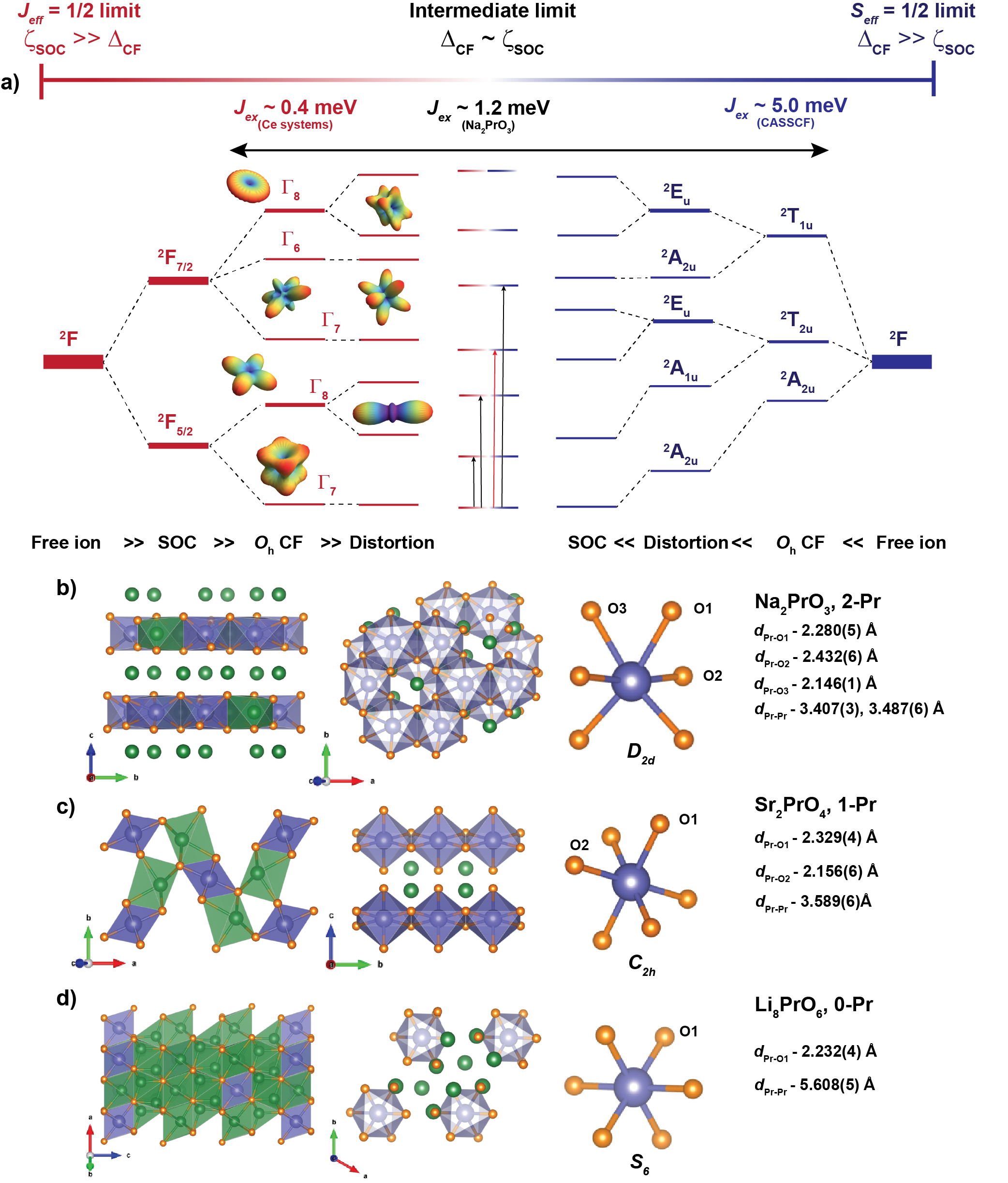}
    \end{center}
    \caption{\label{fig1}
{\bf Competing interactions in Pr$^{4+}$ oxides.} {\bf a}, Hierarchy of single-ion energy scales for Pr$^{4+}$ ions in octahedral oxygen environments starting from a $^2F, (S=1/2, L=3)$ free-ion state. For spin-orbit coupling (SOC) as the dominant energy scale ($J_{\rm eff}\!=\!1/2$ limit, left, brown), the low-symmetry crystal field (CF) lifts the ground-state $^2F_{5/2}$ and excited $^2F_{7/2}$ multiplets into seven Kramers doublets (KDs, with selected radial squared wave-functions represented). For the proximate $O_h$ symmetry, the $\Gamma_7$ doublet is given in the $|J,m_J\rangle$ basis is given by $\Gamma_7^{\pm}\!=\! \sin\theta|\frac{5}{2},\pm\frac{5}{2}\rangle + \cos\theta|\frac{5}{2},\mp\frac{3}{2}\rangle$, where $\sin^2\theta\approx1/6$ (See SI). For CF as the dominant energy scale ($S_{\rm eff}\!=\!1/2$ limit, right, blue), SOC and distortions from $O_h$ symmetry lift the $^2A_{2u}$ ground state and the triply degenerate $^2T_{2u}$ and $^2T_{1u}$ excited states into seven KDs. The ground-state doublet is given in the $|m_l,m_s\rangle$ basis by $|\pm\rangle=A~|\mp3,\pm\frac{1}{2}\rangle-~B~|\mp2,\mp\frac{1}{2}\rangle~+~C~|\pm1,\pm\frac{1}{2}\rangle-~D~|\pm2,\mp\frac{1}{2}\rangle$, where $(A^2/B^2)^{\rm \Gamma_7}\approx2.6$ and $(C^2/D^2)^{\rm \Gamma_7}\approx0.33$ for the $|\Gamma_7 \rangle$ doublet~\cite{willers2015correlation} (See SI). The competition between $\Delta_{\rm CF}$ and $\zeta_{\rm SOC}$ scales in Pr$^{4+}$ yields seven new KD (indicated by the mix brown/blue lines) with magnetic properties that are distinct from the $J_{\rm eff}\!=\!1/2$ and $S_{\rm eff}\!=\!1/2$ limits, such as large magnetic super-exchange achievable by tuning the ligand field.  {\bf b-d}, Crystal structure, magnetic lattice dimensionality, and single-ion environment for the Pr$^{4+}$ oxides studied in this work: Na$_2$PrO$_3$ (\textbf{2-Pr}), Sr$_2$PrO$_4$ (\textbf{1-Pr}), and Li$_8$PrO$_6$ (\textbf{0-Pr}) respectively.      
 	}    
\end{figure}

\begin{figure}[h!]
    \begin{center}
        \includegraphics[width=1.\textwidth]{ 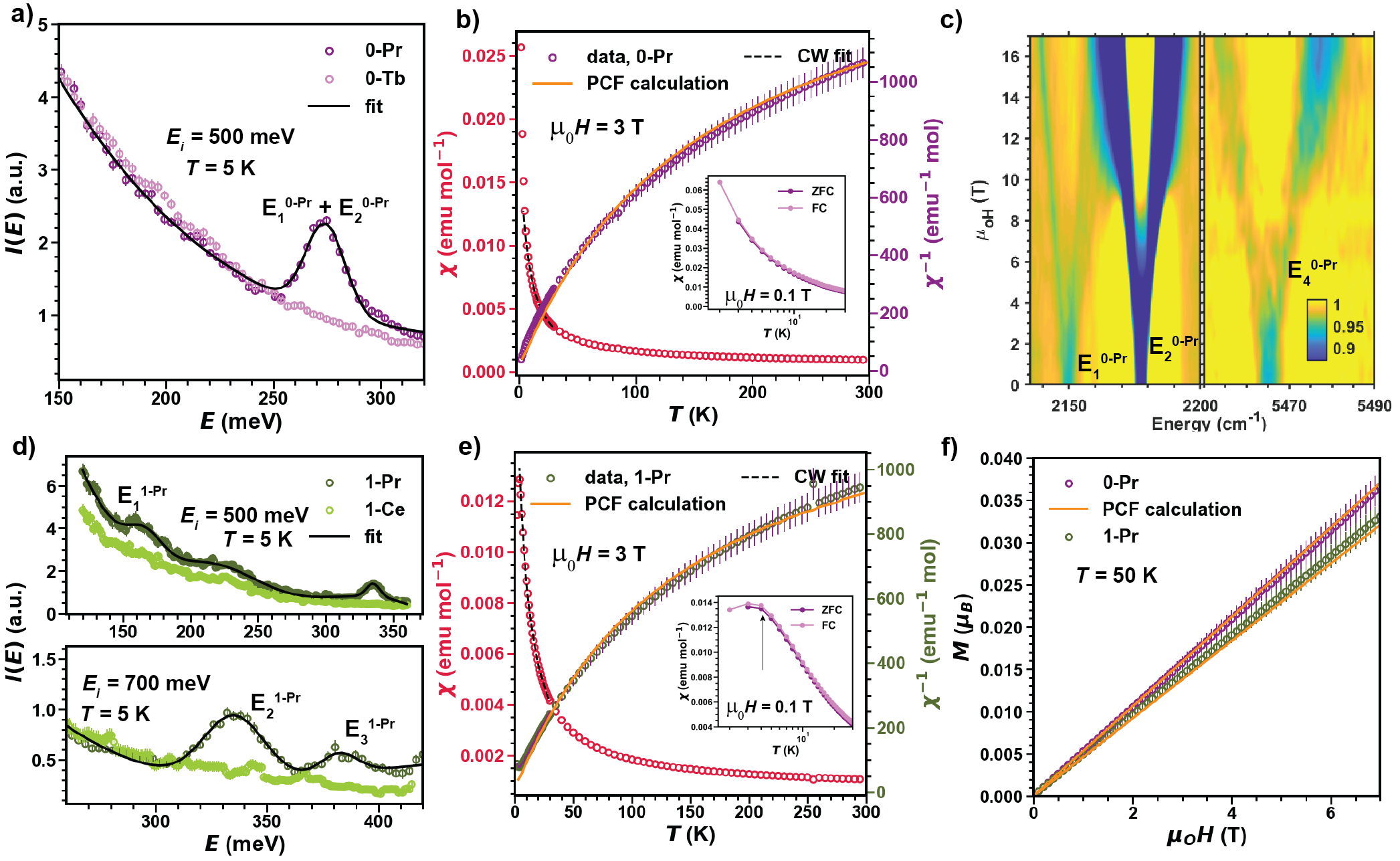}
    \end{center}
    \caption{\label{fig2}
{\bf Crystal field excitations and magnetic properties in Pr$^{4+}$ oxides.} {\bf a,d}, Energy-dependent neutron scattering intensity at low temperature integrated in the range $6<Q<7$ \AA$^{-1}$ for several neutron energies and for \textbf{1-Pr} (Sr$_2$PrO$_4$) and \textbf{0-Pr} (Li$_8$PrO$_6$), respectively. {\bf b,e}, Magnetic susceptibility ($\chi(T)$) and inverse susceptibility ($\chi(T)^{-1}$) data measured under $\mu_0H=3$ T plotted together with CF model and a Curie-Weiss analysis in the temperature range $4<T<40$~K that yields $\Theta_{\rm CW}^{\rm 1-Pr}=-7.52(2)$ K. The CF model calculations were carried out in Stevens operator formalism using the {\scshape Pycrystalfield} package~\cite{scheie2021pycrystalfield} with 14 $|m_l,m_s\rangle$ basis states to account for intermediate coupling. The inset shows macroscopic magnetic behavior under an applied field of $\mu_0H=0.1$~T. \textbf{1-Pr} exhibits an AFM order with a pronounced peak in $\chi(T)$ with no splitting between ZFC and FC. {\bf c}, Normalized IR transmission spectra as a function of applied magnetic field for \textbf{0-Pr}. The blue color indicates the area with intense CEF transitions, while yellow corresponds to the flat line. The experimental data were taken at 5 K and normalized to the average spectra as described in Methods. {\bf f}, Isothermal magnetization $M(H)$ at $T=50$~K for \textbf{0-Pr} and \textbf{1-Pr} plotted together with the CF model. $T=50$~K was chosen so that \textbf{1-Pr} is well above the ordering temperature and free from short-range correlations.   
 	}    
\end{figure}

\begin{figure}[h!]
    \begin{center}
        \includegraphics[width=1.\textwidth]{ 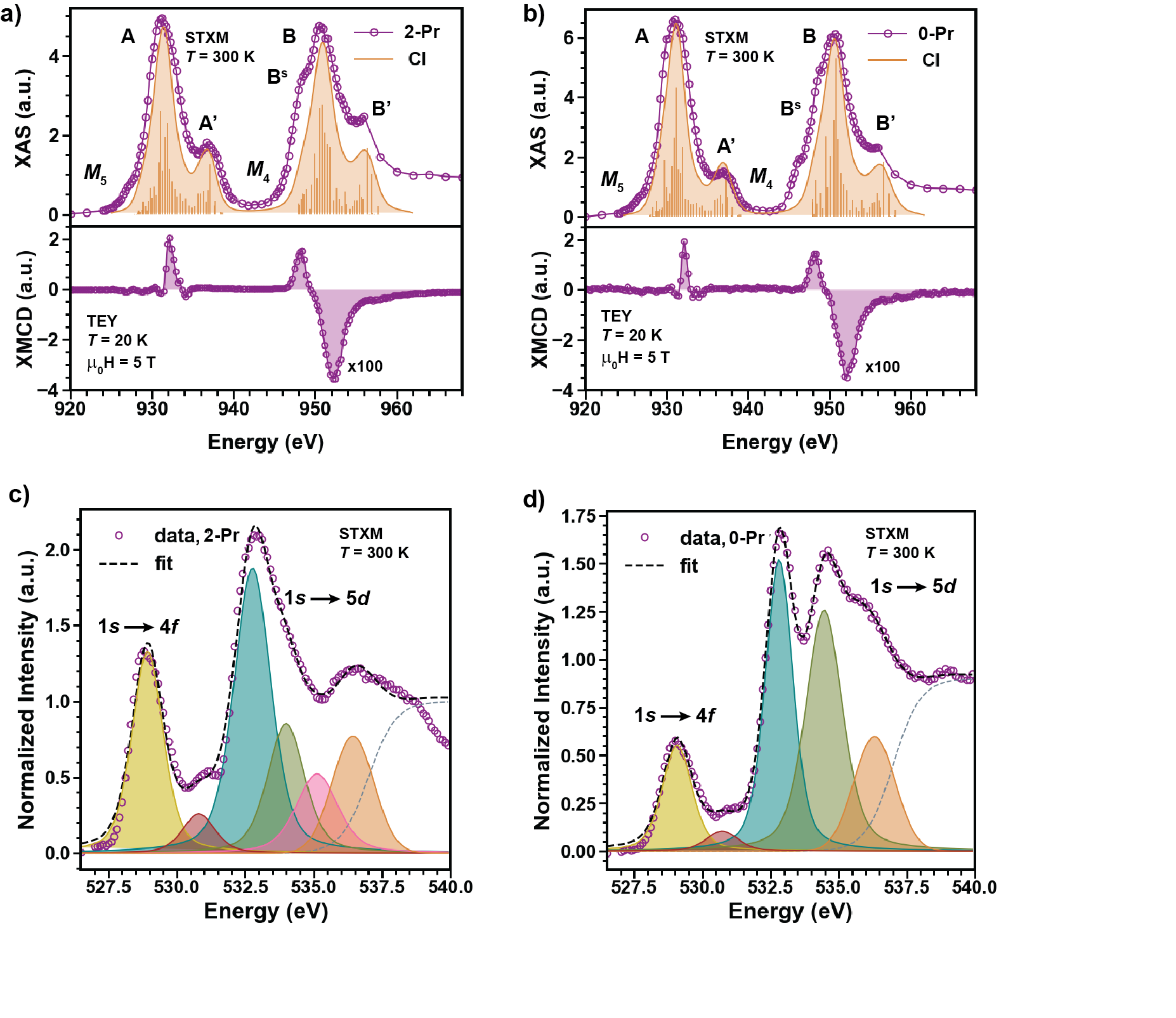}
    \end{center}
    \caption{\label{fig3}
{\bf Fingerprints of Pr-$4f$/O-$2p$ hybridization from X-ray scattering spectra.} {\bf a,b}, Isotropic XAS (top) and XMCD (bottom) spectra at the Pr $M_5$ and $M_4$ edges for \textbf{2-Pr} (left) and \textbf{0-Pr} (right), respectively measured using the Scanning Transmission X-Ray Microscope (STXM) mode (XAS, $\mu_0H=0$ T and $T\!=\!300$ K) and the Total Electron Yield (TEY) mode (XMCD, $\mu_0H=5$ T and $T=20$ K). For the XAS spectra, first-principle calculations (CI under AIM framework) are shown as orange sticks with Gaussian and Lorentzian broadening. For the XMCD spectra, the integration range for the sum-rule analysis is shown as purple shaded region.  {\bf c,d} Isotropic XAS spectra at the oxygen $K$ edge for \textbf{2-Pr} (left) and \textbf{0-Pr} (right), both measured in STXM mode ($T=300$ K). The peak corresponding to Pr-$4f$/O-$2p$ hybridization is shown in yellow. For comparison, reference data on PrO$_2$~\cite{minasian2017quantitative} is shown in Table. S2 and Fig. S4 and S8.}    
\end{figure}

Taken together, the present experiments point at Pr-$4f$/O-$2p$ hybridization as the essential microscopic mechanism behind the unusual electronic and magnetic properties of Pr$^{4+}$ systems. A qualitative understanding of Pr-O bonding is enabled by ab-initio calculations (CASPT2/CASSCF+SOC, see Methods and Suppl. Sec. 4). Considering an isolated [PrO$_6$]$^{8-}$ fragment with perfect $O_h$ symmetry, the $4f$ atomic orbitals (AO) can be easily described in the $|m_l,m_s\rangle$ basis. Here, the CF splitting leads to three spin-free (SF) states: a $^2A_{2u}$ ground-state, and two excited $^2T_{1u}$ and $^2T_{2u}$ triply-degenerate states (Fig~\ref{fig1}a). The $^2A_{2u}$ state has a $\delta$-symmetry with respect to surrounding oxygens and thus remains strictly non-bonding. In contrast, the $^2T_{2u}$ and $^2T_{1u}$ states overlap with oxygen's $2p$ atomic orbitals leading to bonding and anti-bonding molecular orbitals (MO) with respectively  $\pi$ and $\sigma\!+\!\pi$ character about the Pr--O axes (See Fig. S13). When SOC is turned on, Table S9, the ground-state corresponds to the admixture $58\% ^2A_{2u} + 42\% ^2T_{2u}$ (what identifies with the $E_{5/2u}$ term in the $O_h$ double-group symmetry). Departing from $O_h$ symmetry -- as relevant for the PrO$_6$ distorted-octahedra of \textbf{0-Pr} and \textbf{2-Pr} -- lifts the degeneracies of the $^2T_{1u}$ and $^2T_{2u}$ excited states (Tables S10 and S11). However, regardless of symmetry, the ground-state in both \textbf{2-Pr} and \textbf{0-Pr} remains \emph{solely} an admixture of $^2A_{2u}$ and $^2T_{2u}$ states. 

The spectroscopy and thermomagnetic measurements are well explained by this model and calculations. For instance, the calculated $|\langle{L_z}\rangle/\langle{S_z}\rangle|\approx1.8$ (See Table S11) is consistent with the XMCD data, and the small Pr$^{4+}$ magnetic moment can be attributed to self-compensating spin and orbital moments combined with an unusually small $|\langle{L_z}\rangle/\langle{S_z}\rangle|$ that signals a strong reduction of the orbital character in the original $\Gamma_7$ ground-state doublet. This framework naturally explains the O $K$ edge spectra of \textbf{2-Pr} and \textbf{0-Pr} through ligand holes induced by the formation of hybridized $T_{1u}2p_{\sigma} + T_{2u}2p_{\sigma+\pi}$ states. The model also explains why the largest hybridization is observed for PrO$_2$: the eight-, rather than six-, oxygen coordination environment allows the Pr 4f $a_{2u}$ orbital to covalently interact with the O $2p$ orbitals with $\sigma$ symmetry, thereby exhibiting enhanced $4f$-$2p$ hybridization\cite{minasian2017quantitative}. The difference in $4f$-$2p$ hybridization between \textbf{2-Pr} and \textbf{0-Pr} likely comes from different point-group symmetries for the PrO$_6$ unit and the overall symmetry of the material. It is clear that $4f$-$2p$ hybridization can strongly influence single-ion energy scales such as the CF - this phenomenon is directly analogous to the behaviour of  $d$-block metals. 

\begin{figure}[h!]
    \begin{center}
        \includegraphics[width=1.\textwidth]{ 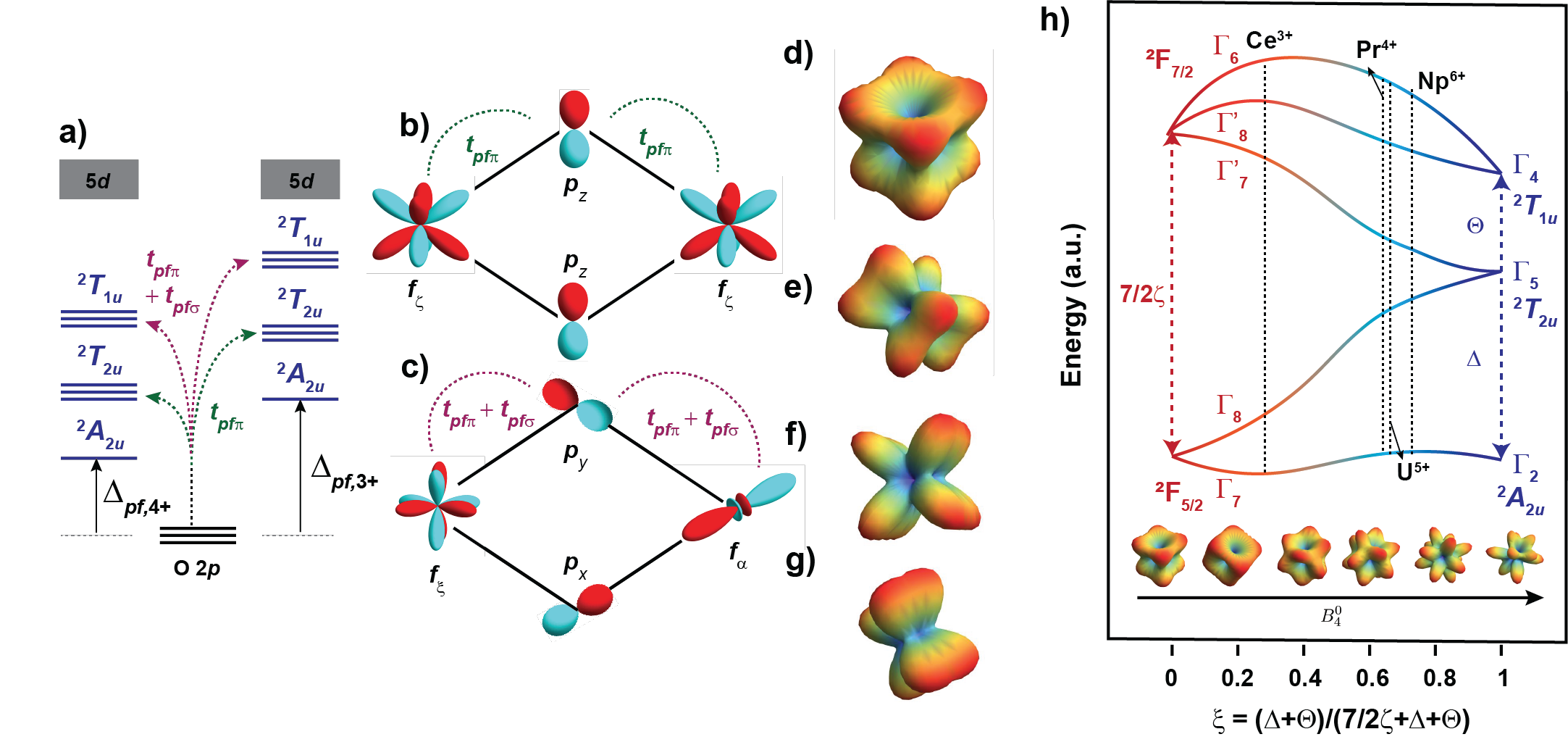}
    \end{center}
    \caption{\label{fig4}
{\bf Microscopic origin of anomalous properties of Pr$^{4+}$ and a universal model for $f^1$ single-ions.} {\bf a}, Schematic of $p$ and $4f$ energy levels for Pr$^{4+}$ and Ce$^{3+}$. $t_{pf\pi}$ ($t_{pf(\pi+\sigma)}$) is the hopping integral between $p$ and $^2T_{2u}$ ($^2T_{1u}$) orbitals. The corresponding $pf$ charge transfer gap is indicated with $\Delta_{pf,4+} < \Delta_{pf,3+}$. {\bf b}, Sketch of the hopping processes between occupied $f_{\zeta}$ orbitals mediated by the $\pi$ interacting $2p$ orbitals analogous to $t_{2g} - p - t_{2g}$ hopping in $d^5$ systems. {\bf c}, Sketch of the hopping processes between occupied $f_{\zeta}$ and unoccupied $f_{\alpha}$ orbitals mediated by the $\pi + \sigma$ interacting $2p$ orbitals analogous to $t_{2g} - p - e_{g}$ hopping in $d^5$ systems. {\bf d-g}, Probability density of the ground state KD in ideal $\Gamma_7$, \textbf{2-Pr}, \textbf{1-Pr}, and \textbf{0-Pr}, respectively and shows the impact of mixing excited states in to the original $\Gamma_7$ doublet. {\bf h}, Schematic of the splitting of $f$ orbitals as a function of CF ($\Delta$ and $\theta$) relative to SOC ($\zeta$). The value of $\xi$ for Pr$^{4+}$ was calculated from \textbf{0-Pr}, and the values for Ce$^{3+}$, U$^{5+}$, and Np$^{6+}$ were obtained from\cite{lukens2013quantifying}. Using this as a universal model for $f^1$ ions, Pr$^{4+}$ is categorized together with the actinides, where the traditional Ln$^{3+}$ picture breaks down. The figure also shows the evolution of the shape of the $\Gamma_7$ KD as a function of $B^0_4$ in the $\mathcal{\hat{H}}_{CF}^{O_h} = B^0_4\hat{O}^0_4+ B^4_4\hat{O}^4_4$, where $B^4_4 = 5B^0_4$. Increasing $B^0_4$ from $\sim0$ ($\zeta_{\rm SOC} >> \Delta_{\rm CF}$) yielding a almost perfect $\Gamma_7$ KD (left most figure) to $\sim2000$ ($\zeta_{\rm SOC} << \Delta_{\rm CF}$, right most figure) and the resulant drastic changes of the nature of the KD. The original nature of the $\Gamma_7$ KD is retained until the eigen value of the $\Gamma_8\approx75$ meV where $J_{\rm eff}\!=\!1/2$ limit still applies.       
 	}    
\end{figure}

Beyond single-ion properties, Pr-$4f$/O-$2p$ hybridization leads to unusually large two-ion magnetic exchange interactions. For example, $J=1.2$~meV has been reported by some of us on \textbf{2-Pr} \cite{daum2021collective}; a value 2.5 times larger than the typical scale of $J\approx0.4$~meV observed for $4f^1$ or $4f^{13}$ systems such as KCeO$_2$ and NaYbO$_2$~\cite{bordelon2019field,bordelon2021magnetic}. The Weiss constant of \textbf{1-Pr} of around $|\Theta_{\rm CW}|=7$~K is also large, especially considering the quasi-1D nature of this system. MO theory can be used to understand these exchange interactions, as shown in Fig~\ref{fig4}a. In the charge transfer limit ($U_{ff}\gg\Delta E_g$), the nearest-neighbor exchange interaction scales as $t_{pf}^4/\Delta E_{pf}^3$, where $t_{pf}$ is the hopping integral between $4f$ and $2p$ orbitals, and $\Delta E_{pf}$ is their energy difference (i.e. the ligand to metal charge transfer (LMCT)). The enhancement of magnetic exchange in Ln$^{4+}$ compounds is likely primarily driven by the reduction of the charge transfer energy $\Delta E_{pf}$ , as evidenced by \textit{e.g.} calculations of [CeCl$_6$]$^{3-}$ ($>5$ eV) and [CeCl$_6$]$^{2-}$ ($\sim 3.3$ eV)~\cite{loble2015covalency}. Large $t_{pf}$ hopping and small $\Delta_{pf}$ implies a large ligand-hole character, consistent with our O $K$-edge spectra. First principles calculations on a binuclear [Pr$_2$O$_{10}$]$^{12-}$ embedded cluster model for \textbf{2-Pr} (see Suppl. Sec. 4) quantitatively confirm this picture. In the $S_{\rm eff}\!=\!1/2$ limit, the spin-singlet minus spin-triplet energy, which identifies with the Heisenberg exchange interaction, yields $J=4.2$~meV (See Fig. S14). Upon including SOC, $J$ reduces to $\approx1.5$ meV, in good agreement with experimental results for \textbf{2-Pr}. Therefore exchange interactions in Pr$^{4+}$ materials may change by an order of magnitude (0.3~meV to 4.2~meV) under changes of the ligand field (Fig~\ref{fig1}a). Similar effects have been observed in high-valent actinides including U$^{5+}$ and Np$^{6+}$\cite{magnani2010superexchange}. Inspecting hopping pathways is also informative to comment on the Kitaev (AFM/FM) interactions proposed for \textbf{2-Pr} through the $T_{2u}-p-T_{1u}$ pathway (Fig~\ref{fig4}b) analogous to the $t_{2g}-p-e_g$ process in $d^5$ systems~\cite{jang2019antiferromagnetic}. While this contribution is small in $d^5$ systems due to the large $t_{2g}$ to $e_g$ separation, it is proposed to be larger in $4f^1$-systems owing to the small CF energy scale~\cite{jang2019antiferromagnetic}. However, as we have demonstrated, Pr$^{4+}$ systems exhibits a very large CF splitting (or in other words, the $J_{\rm eff}\!=\!1/2$ limit is not adequate), making the $T_{2u}-p-T_{1u}$ pathway energetically less favorable~\cite{liu2022exchange} than the $T_{2u}-p-T_{2u}$ pathway (Fig~\ref{fig4}c) responsible for the large Heisenberg AFM interaction.

Finally, and very importantly, the competition between CF and SOC energy scale in Pr$^{4+}$ systems resembles that in high-valent actinide systems such as U$^{5+}$ and Np$^{6+}$ for which a CF energy scale as large as $\approx800$ meV is possible~\cite{lukens2013quantifying,eisenstein1960theory}. The chemistry and physics of high-valent actinides is further complicated by an extra competition between Coulomb repulsion, CF, and SOC \cite{moore2003failure,shim2007fluctuating} leading to the dual nature of $5f$ electrons\cite{booth2012multiconfigurational}. In order to develop a universal description for $f^1$ single-ions, we argue that Pr$^{4+}$ systems can facilitate the study of the delicate balance of various competing interactions in absence of competing Coulomb repulsion. To showcase that idea, we use the model established by Eisentein and Pryce~\cite{eisenstein1960theory} where, the CF transitions for a $f^1$ ion in an ideal $O_h$ symmetry can be written as a function of two CF parameters $\Delta$ and $\theta$, and $\zeta_{\rm SOC}$, as shown in Fig~\ref{fig4}f. Using this framework and our experimentally determined values for \textbf{0-Pr}, we calculate the parameter $\xi=\frac{\Delta+\theta}{7/2\zeta+\Delta+\theta} \approx 0.62$ for Pr$^{4+}$. When compared with other $f^1$ ions including Ce$^{3+}$, U$^{5+}$, and Np$^{6+}$ (see Fig~\ref{fig4}h and Refs. \cite{lukens2013quantifying,edelstein2008magnetic}), it is evident that Pr$^{4+}$ lies closer to U$^{5+}$ than Ce$^{3+}$. Qualitatively, the observed trend can be generalized by a simple $f$-orbital bonding picture that puts hybridization with the ligands $np$-orbitals as the key microscopic phenomena leading to an enhanced CF energy scale.

In summary, our work elucidated the mechanisms behind the anomalously large CF energy scale in Pr$^{4+}$ systems and discussed how exotic magnetic and electronic properties emerge as a result. The covalent character of the Pr--O bond plays a key role in determining the single-ion and macroscopic physics in Pr$^{4+}$ compounds, similar to familiar systems such cuprates and nickelates. It is in sharp contrast to Ln$^{3+}$ systems for which, conventionally, the metal-ligand bond is described using an ionic picture.  Our results offer novel strategies to design and control quantum materials by tuning the Pr--O covalency through site symmetry and ligand identity. The inadequacy of the  $J_{\rm eff}\!=\!1/2$ limit shows us how to change the fabric of spin-orbit entangled single-ion states to stabilize exotic exchange Hamiltonians or to develop universal models to understand the physics of high-valent actinides. Pr$^{4+}$-based systems offer the rare possibility to tune the hierarchy of single-ion energy scales, as well as the $p$ and $f$ hole density, which may be harnessed to design new correlated phenomena in quantum materials.

\clearpage

\clearpage

\newpage

\newpage

\newpage
\clearpage

\clearpage
\section*{References}
\renewcommand\refname{\vskip -1cm}

\clearpage
\section*{Acknowledgments}

We are thankful to Dr. Harry Lane for insightful discussions. The work of A.R. and H.S.L. at Georgia Tech was supported by the Beckman Foundation as part of a Beckman Young Investigator Award to H.S.L. The work of J.K. and M.M. at Georgia Tech was supported by the National Science Foundation through Grant No. NSF-DMR-1750186. The work of Z.J. at Georgia Tech was supported by the U.S. Department of Energy through Grant No. DE-FG02-07ER46451. Some of this work was performed in part at the Materials Characterization Facility at Georgia Tech which is jointly supported by the GT Institute for Materials and the Institute for Electronics and Nanotechnology, and is a member of the National Nanotechnology Coordinated Infrastructure supported by the National Science Foundation under Grant No.~ECCS-2025462. This research used resources at the Spallation Neutron Source, a DOE Office of Science User Facility operated by the Oak Ridge National Laboratory. Use of the Advanced Photon Source at Argonne National Laboratory was supported by the U.S. Department of Energy, Office of Science, Office of Basic Energy Sciences, under Contract DE-AC02-06CH11357. The infrared measurements were performed at the National High Magnetic Field Laboratory, which is supported by the National Science Foundation Cooperative Agreement No. DMR-1644779 and the State of Florida. The work of D.C.S. and J.A. at the University at Buffalo was supported by the U.S. Department of Energy, Office of Basic Energy Sciences, Heavy Element Chemistry program, under grant DESC0001136. D.C.S. and J.A. thank the Center for Computational Research (CCR) at the University at Buffalo for providing computational resources. D.C.S. received research funding from the European Union's Horizon 2020 Research and Program under Marie Sklodowska-Curie Grant Agreement No. 899546. D.C.S. acknowledges infrastructure support provided through the RECENT AIR grant agreement MySMIS no. 127324. Work of J.A.B. and S.G.M at LBNL was supported by the Director, Office of Science, Office of Basic Energy Sciences, Division of Chemical Sciences, Geosciences, and Biosciences of the U.S. Department of Energy (DOE) at LBNL under Contract No. DE-AC02-05CH11231. STXM research described in this paper was performed at the Canadian Light Source, which is supported by the Canada Foundation for Innovation, Natural Sciences and Engineering Research Council of Canada, the University of Saskatchewan, the Government of Saskatchewan, Western Economic Diversification Canada, the National Research Council Canada, and the Canadian Institutes of Health Research\\

\noindent {\bf Authors contributions}: A.R., M.M. and H.S.L. conceived the project which was led by A.R. and H.S.L. A.R. synthesized the sample at Georgia Tech. A.R., J.K., A.I.K., and M.M. performed the neutron-scattering measurements. A.R. and J.K. analyzed the neutron scattering data. A.R., M.M. and J.K. carried out thermo-magnetic measurements. M.O. and Z.J. performed the IR measurements and analyzed the data. J.A.B. and S.G.M. performed STXM XAS measurements and normalized the data. A.R., J.W.F., and H.S.L. measured XMCD data. A.R. analyzed the X-ray scattering data. D.C.S. and J.A. carried out the theoretical calculations and accompanying analyses. A.R., M.M. and H.S.L. wrote the manuscript with input from all authors.\\

\noindent {\bf Competing interests}: The authors declare not competing interests.\\

\noindent {\bf Corresponding author}: Correspondence to Henry S. La Pierre.\\

\noindent {\bf Data and materials availability}: All data in the manuscript or the supplementary materials is available upon email requests to Henry S. La Pierre.\\

\clearpage
\noindent\subsection*{}
       \renewcommand\refname{References}
       \renewcommand{\thesection}{\arabic{section}}
       \renewcommand{\thesubsection}{\thesection.\arabic{subsection}}
        \setcounter{equation}{0}
        \makeatletter 
        \def\tagform@#1{\maketag@@@{(S\ignorespaces#1\unskip\@@italiccorr)}}
        \makeatother
        \setcounter{figure}{0}
        \makeatletter
        \makeatletter \renewcommand{\fnum@figure}
        {\figurename~S\thefigure}
        \makeatother
        \setcounter{table}{0}
        \makeatletter
        \makeatletter \renewcommand{\fnum@table}
        {\tablename~S\thetable}
        \makeatother

\newcommand*\mycaption[2]{\caption[#1]{#1#2}}


\section*{Methods}

    \subsection*{Material synthesis.}
Na$_2$LnO$_3$ (1-Ln, Ln = Ce, Pr) were synthesized using established procedures\cite{ramanathan2021plane}. Polycrystalline powder samples of Sr$_2$LnO$_4$ (2-Ln, Ln = Ce, Pr) and Li$_8$LnO$_6$ (3-Ln, Ln = Pr, Tb) were synthesized using traditional solid-state methods. The samples were fired under a flow of O$_2$ in a tube furnace. The firing was performed at $1100\degree$ C for $24$ h. The samples were taken out of the quartz tubes in air and placed into the antechamber of the glovebox as quickly as possible in order to minimize contact with the ambient atmosphere (See SI for details). 
    \subsection*{Experimental Characterizations.}
        \subsubsection*{PXRD.}
Laboratory powder X-ray diffraction (PXRD) was collected on a PANalytical X’Pert PRO Alpha-1 diffractometer with Cu $K\alpha$ source in reflection geometry equipped with a fixed divergence slit of $1/8"$, a convergence slit of $1/4"$ and a working radius of $240$ mm. The samples were homogenized by finely grinding them inside the glove box using an agate mortar for about $\sim15$ min. To avoid the exposure of sample to atmospheric air, PANalytical domed sample holder equipped with stainless steel base and a polycarbonate dome with a $70\%$ X-ray transmission. A $2\theta$ range of $5-\:85\degree$ was used with a scan speed of $5$ s and a step size of $0.1$. 
        \subsubsection*{Physical property measurements.}
The d.c. magnetic susceptibility measurements and isothermal magnetization measurements were using a Quantum Design MPMS3. The sample was sealed in a plastic capsule on a low-background brass holder.
        \subsubsection*{Neutron scattering measurements.}
Inelastic neutron scattering measurements were carried out on $\sim8$ g of polycrystalline samples of \textbf{1-Pr}, \textbf{1-Ce}, \textbf{0-Pr}, and \textbf{0-Tb} on the time-of-flight fine-resolution Fermi chopper spectrometer SEQUOIA, at the Spallation Neutron Source, ORNL~\citeMethods{granroth2010sequoia,stone2014comparison}. The powder samples were enclosed in a standard 15-mm diameter cylindrical aluminum cans under one atmosphere of $^4$He at room temperature. All four samples were cooled using a closed-cycle refrigerator reaching a base temperature of $T = 5$ K. Measurements were carried out using incident neutron energies $E_i=300, 500, 700$ meV at $T=5$~ K. Background and sample holder contributions were measured using empty can measurements. The lattice phonon contributions for \textbf{0-Pr} and \textbf{1-Pr} were subtracted by measuring the analogous \textbf{0-Tb} and \textbf{1-Ce}, respectively. 
        \subsubsection*{STXM O \texorpdfstring{$K$}~~edge XAS.}
STXM methodology was similar to that discussed previously \cite{minasian2017quantitative}. In an argon-filled glovebox, samples for STXM measurement were prepared by pulverizing the powder compounds and transferring particles to Si$_3$N$_4$ windows. Second windows were placed over the samples to sandwich the particles, and the windows were sealed together with Hardman Double/Bubble epoxy.  Single-energy images and O $K$-edge XAS spectra were acquired using the STXM instrument at the Canadian Light Source (CLS) spectromicroscopy beamline 10ID-1, operating in decay mode ($250$ to $150$ mA, in a $\sim0.5$ atm He-filled chamber) at a base temperature of $T = 300$ K. The beamline uses photons from an elliptically polarizing undulator that delivers photons in the 130 to 2700 eV energy range to an entrance slit-less plane-grating monochromator. The maximum energy resolution $E/\Delta E$ was previously determined to be better than 7500, which is consistent with the observed standard deviation for spectral transitions of $\pm~0.1$ eV determined from the comparison of spectral features over multiple particles and beam runs. For these measurements, the X-ray beam was focused with a zone plate onto the sample, and the transmitted light was detected. The spot size and spectral resolution were determined from the characteristics of the 35 nm zone plate. Images at a single energy were obtained by raster-scanning the sample and collecting transmitted monochromatic light as a function of the sample position. Spectra at particular regions of interest on the sample image were extracted from the “stack”, which is a collection of images recorded at multiple, closely spaced photon energies across the absorption edge. Dwell times used to acquire an image at a single photon energy were 2 ms per pixel and spectra were obtained using circularly polarized radiation. The incident beam intensity was measured through the sample-free region of the Si$_3$N$_4$ windows. In order to ensure that the spectra were in the linear regime of Beer–Lambert's law, particles with an absorption of less than $1.5$ OD were used. High-quality spectra were obtained by averaging measurements from multiple independent particles, samples, and beam runs.  
        \subsubsection*{STXM Pr \texorpdfstring{$M_{5,4}$}~~edge XAS.}
Measurements at the Pr $M_{5,4}$-edges were conducted using the STXM instrument at the Canadian Light Source (CLS) spectromicroscopy beamline 10ID-1, operating in top-off mode ($250$ mA, in a $\sim0.5$ atm He-filled chamber) at a base temperature of $T = 300$ K. The sample preparation and data acquisition methodology is the same as described above for the O $K$-edge measurements. 
        \subsubsection*{Pr \texorpdfstring{$M_{5,4}$}~~edge XMCD.}
The XAS and XMCD measurements at Pr $M_{5,4}$-edges were conducted at beamline 4-ID-C of the Advanced Photon Source located at Argonne National Laboratory. XAS and XMCD spectra were collected simultaneously using total electron yield (TEY) and total fluorescence yield (TFY), respectively, with circularly polarized X-rays in a near normal ($80\degree$) configuration using a cryostat reaching a base temperature of $T=20$ K. The applied field was along the beam direction and it defines the positive $Z$ direction. The data was obtained at both zero field and an applied field of $\mu_OH = \pm5$ T. The XMCD spectra were obtained point by point by subtracting right from left circular polarized XAS data. Measurements were taken for both positive and negative applied field directions and then a difference of these two spectra XMCD = $\dfrac{1}{2}$ [XMCD($H_z>0$) - XMCD($H_z<0$)] was taken to eliminate polarization-dependent systematic errors. The TFY XAS data is identical to the STXM data described above. However, the TFY XMCD signal is weak and distorted by self-absorption effects. The TEY XAS data is similar to STXM data as well, except the high energy satellite peaks at both $M_{5,4}$ edges are weak and not as pronounced. Furthermore, the low energy shoulder at the $M_4$ edge is more pronounced in TEY XAS than in both TFY and STXM. For discussions in the main text regarding $M_{5,4}$ edge isotropic XAS spectra, only the STXM data is discussed as it minimizes error due to self-absorption, saturation, and surface-contamination. However for our discussions with XMCD, we use the data collected in TEY mode. As noted in Fig xx, the isotropic XAS was in the top panel and is measured in STXM mode, while the XMCD at the bottom panel and is measured in TEY.  
        \subsubsection*{Infrared magnetospectroscopy.}
Broadband IR measurements were performed in the Voigt transmission configuration using a Bruker 80v Fourier-transform IR spectrometer. The incident IR light from a globar source was guided to the top of the probe inside an evacuated beamline and then delivered to the bottom of the probe through brass light pipes. The sample was located in the middle of two confocal $90^{o}$ off-axis parabolic mirrors mounted at bottom of the probe. While the first mirror focuses the IR radiation on the sample, the second mirror collimates the radiation to the short light pipe with a 4K composite Si bolometer at the end. About 25 mg of the powder sample was mixed with KBr inside a glovebox in 1:1 ratio. The resulting mixture was pressed in to 3mm pellets and was secured by a thin polypropylene adhesive film and mounted on the brass plate with a clear aperture 3mm. The sample was placed at the center of a $\mu_0H=17.5$~T vertical bore superconducting magnet in a helium exchange gas environment, providing the sample temperature of about 5.5 K. IR transmission spectra were collected for 3 min at a fixed magnetic field, changing with 1 T step. All spectra obtained at different magnetic fields were normalized to the same reference spectrum, which is their mean, computed after removing the outlier points at each frequency. Such normalization flattens those spectral features independent of magnetic fields and highlights those absorption peaks that shift as the magnetic field rises.

    \subsection*{Crystal field modeling of inelastic neutron scattering (INS).}
CF modeling was carried out using the truncated CF Hamiltonian $\mathcal{\hat{H}}_{\rm CF} = B^0_2 \hat{O}^0_2 + B^0_4\hat{O}^0_4+ B^4_4\hat{O}^4_4 + B^0_6\hat{O}^0_6 + B^4_6\hat{O}^4_6$ where $B^m_n$ are the second, fourth, and sixth order terms and $\hat{O}^n_m$ are the corresponding Stevens operator equivalents\citeMethods{stevens1952matrix} for all three materials studied here. Although the true symmetries of Pr$^{4+}$ in each system requires more parameters based on point-group symmetry, any mixing induced by these parameters would not induce any further loss of degeneracy and hence we choose to parameterize their effects  $B^0_n$ and $B^4_n$ parameters. This approach was taken in order to minimize over-parametrization while fitting to experimental data. All Hamiltonian diagonalizations were performed using the {\scshape Pycrystalfield} package~\cite{scheie2021pycrystalfield}. Fitting was carried out to a combination of eigen-energies extracted experimentally from INS and IRMS and to the temperature-dependent susceptibility data over $T>40$ K in order to avoid short-range correlations present at lower temperatures. The final fit results are provided in Suppl. Tab. S3. The CF models were validated by calculating the isothermal magnetization at $T=50$~K. The model calculation of $g$ values for the ground state wavefunction was compared to experimentally determined values from Curie-Weiss fits and first-principles calculations. See Suppl. Sec. 3 for a detailed description of the fitting procedure.  

The $\Gamma_7$ KD in the $\Delta_{\rm CF} << \zeta_{\rm SOC}$ limit is written in the $|J,m_J\rangle$ basis as $\sin \theta|\frac{5}{2},\pm\frac{5}{2}\rangle +\cos\theta|\frac{5}{2},\mp\frac{3}{2}\rangle$, where $\sin \theta^2\sim1/6$. The same $\Gamma_7$ KD can be written in the $|m_l,m_s\rangle$ basis as $A~|\mp3,\pm\frac{1}{2}\rangle-~B~|\mp2,\mp\frac{1}{2}\rangle~+~C~|\pm1,\pm\frac{1}{2}\rangle-~D~|\pm2,\mp\frac{1}{2}\rangle$, where $A=0.352$, $B=0.215$,$C=0.454$,$D=0.79$, yielding $\alpha=\frac{A^2+B^2}{C^2+D^2}\approx0.18$. 
The first two components of the $\Gamma_7$ KD in $|m_l,m_s\rangle$ ($m_l=\mp3,\mp2$) map onto a linear combination of the $|\frac{5}{2},\pm\frac{5}{2}\rangle, |\frac{7}{2},\pm\frac{5}{2}\rangle$, states in $|J,m_J\rangle$ basis, while the last components ($m_l=\pm1,\pm2$) map onto $|\frac{5}{2},\pm\frac{3}{2}\rangle, |\frac{7}{2},\pm\frac{3}{2}\rangle$ states. For the $\Gamma_7$ KD, given that $\Delta_{\rm CF} << \zeta_{\rm SOC}$, the contributions from the $J=\frac{7}{2}$ SOC manifold are negligible. As  $\Delta_{\rm CF} \sim \zeta_{\rm SOC}$, non-negligible contributions from the $J=\frac{7}{2}$ SOC manifold enter the ground-state wavefunction making it impossible to deconvolute the individual contributions from each SOC manifold. Therefore, a better description of mixing can be obtained by looking at the ratios $\frac{A^2}{B^2}$ and $\frac{C^2}{D^2}$. Within this framework, irrespective of the symmetry at the metal center, for a six-coordinate system, the ground state wavefunction is always a linear combination of $m_l=\pm1,\pm2,\mp3$ states. This derives from the $^2A_{2u} + ^2T_{2u}$ symmetry (in $O_h$ notation) as described in the main text and predicted by first-principles calculations. Introduction of intermediate coupling, changes only the relative mixtures of $m_l=\pm1,\pm2,\mp3$ states and does not introduce any new admixture into the ground state wavefunction. The relative change in mixture can be viewed as introducing $|\frac{7}{2},\pm\frac{5}{2}\rangle$ and $\frac{7}{2},\pm\frac{3}{2}\rangle$ states and increasing the amount of $|J,\pm\frac{5}{2}\rangle$ character in the ground state. This is clearly evident in the toy model established in Supplementary Section. 3.    

    \subsection*{Multiplet modeling of x-ray absorption spectroscopy (XAS).}
Multiplet calculations were implemented using the original code written by Cowan\cite{cowan1981theory} and further developed by de Groot based on AIM. The multi-electron configuration in the ground and the final states was implemented using a charge-transfer methodology analogous to nickelates and cuprates. For all calculations, a Gaussian broadening of $0.45$ eV was applied to account for instrumental broadening and Lorentzian broadening of $0.3$ and $0.6$ eV were applied to the $M_5$ and $M_4$ edges, respectively. The model parameters had the following values for both 1-Pr and 3-Pr: $U_{ff}\sim14.1$ eV, $U_{fc}\sim8.5$ eV, $\zeta_{SOC}^{4f}\sim0.12$ eV, and $\zeta_{SOC}^{3d}\sim7.1$ eV where $U_{ff}$ and $U_{fc}$ are the $4f-4f$ Coulomb interaction and the core-hole potential acting on the $4f$ electron, respectively. In the limit of vanishing $V\rightarrow0$, the difference between the two configurations in the ground state was $\Delta E_{g}=\epsilon_f-\epsilon_n=2.0$ eV (1-Pr) and $3.0$ eV (3-Pr), and $\Delta E_{f}=\epsilon_f-\epsilon_n+U_{ff}-U_{fc}=0.5$ eV (1-Pr) and $1.5$ eV (3-Pr), where $\epsilon_f$ and $\epsilon_n$ are one-electron energies of Pr $4f$ and O $2p$ levels and $V$ is the hybridization energy between atomic like localized $4f$ states and delocalized O $2p$ states which determines the mixing between the multi-electron configurations. Hybridization energy in the ground state ($V_g$) was set to $1.4$ eV (1-Pr and 3-Pr) and final state ($V_f$) was set to $1.4$ eV (1-Pr) and $1.8$ eV (3-Pr).   
        
    \subsection*{First-principles calculations.}
Without symmetry restrictions, single-point wavefunction theory (WFT) calculations were performed within a commonly applied two-step spin-orbit coupled configuration interaction formalism using OpenMolcas.\citeMethods{Autschbach:2020n} WFT calculations (See SI for more details) were performed on isolated PrO$_6^{8-}$ ions and on embedded cluster models for Li$_8$PrO$_6$ and Na$_2$PrO$_3$. Additional calculations were performed with a binuclear Pr$_2$O$_{10}^{12-}$ embedded cluster model of Na$_2$PrO$_3$. All geometries were extracted from the crystal structures and used without further optimization. A 40 \AA~sphere of atoms was generated from the crystal structure. An outer 32 \AA~sphere contained Pr$^{4+}$, O$^{2-}$, Na$^{+}$/Li$^{+}$ embedding point charges; the inner 8 \AA ~sphere contained the PrO$_6^{8-}$ ion treated quantum-mechanically surrounded by Pr$^{4+}$, O$^{2-}$, Na$^{+}$/Li$^{+}$ pseudocharges represented by ab-initio model potentials (AIMPs). All atoms treated quantum mechanically were modeled with all-electron atomic natural orbital relativistically contracted basis sets of valence triple-$\zeta$ quality (ANO-RCC-VTZP).

\renewcommand\refname{\vskip -1cm}

\end{document}



\clearpage

\noindent\subsection*{}
       \renewcommand\refname{References}
       \renewcommand{\thesection}{\arabic{section}}
       \renewcommand{\thesubsection}{\thesection.\arabic{subsection}}
        \setcounter{equation}{0}
        \makeatletter 
        \def\tagform@#1{\maketag@@@{(S\ignorespaces#1\unskip\@@italiccorr)}}
        \makeatother
        \setcounter{figure}{0}
        \makeatletter
        \makeatletter \renewcommand{\fnum@figure}
        {\figurename~S\thefigure}
        \makeatother
        \setcounter{table}{0}
        \makeatletter
        \makeatletter \renewcommand{\fnum@table}
        {\tablename~S\thetable}
        \makeatother

\newcommand\mycaption[2]{\caption[#1]{#1#2}}
\setcounter{secnumdepth}{4}
\renewcommand{\contentsname}{Supplementary Materials}
\tableofcontents

\clearpage

\section{Material Synthesis.}\label{sec1}
All reagents were handled in a N$_2$ filled glove box (Vigor) with O$_2 < 0.1 $ ppm and H$_2$O $< 0.1$ ppm. Na$_2$O (Alfa Aesar), SrCO$_3$ (99.5\%, Alfa Aesar), Pr$_6$O$_{11}$($\geq99.5\%$, Alfa Aesar), Tb$_4$O$_7$ ($\geq99.998\%$, Alfa Aesar), CeO$_2$ ($\geq99.9\%$, Alfa Aesar), and Li$_2$O ($\geq99.9\%$, Alfa Aesar) were used as starting materials. The metal oxides and SrCO$_3$ powders were dried by heating to $500\degree$ C for $~12$ h with a heating rate of $10\degree$ C/min in a box furnace (using alumina crucibles) under ambient atmosphere. The reagents were then cooled with the furnace off to $\sim120\degree$ C, and then cooled to room temperature in the antechamber of the glovebox under vacuum. These dried reagents were stored in amber bottles in the glove box. An MTI-KSL-1100X-S-Ul-LD furnace was used. All crucibles were purchased from MTI.
    \subsection{Synthesis of 2-Ln (Ln = Ce, Pr).}\label{sec1.1}
Na$_2$LnO$_3$ (2-Ln, Ln = Ce, Pr) was synthesized following prior published work\cite{ramanathan2021plane}. 
    \subsection{Synthesis of 1-Ln (Ln = Ce, Pr).}\label{sec1.2}
Polycrystalline powder samples of Sr$_2$LnO$_4$ (1-Ln, Ln = Ce, Pr) were synthesized using traditional solid-state methods by intimately mixing SrCO$_3$ and Pr$_6$O$_{11}$ (CeO$_2$) in molar ratio 2.0:1 (Sr:Ln), using an agate mortar inside the glove box. The powder mixtures were pressed in to $15$ mm diameter pellets outside the glovebox. The samples were fired under a flow of O$_2$ in tube furnace (quartz tubes with a diameter of $55$ mm was used). The O$_2$ flow was controlled using a regulator set to $2$ psig and an oil bubbler at the end of the line to $\sim1$ bubble every 2-3 sec. The pellets were placed on alumina boats and placed at the center of the quartz tube (lining up with the center of the heating zone in the furnace). The line was then purged with O$_2$ for $\sim5$ min. The firing was performed at $1100\degree$ C for $24$ h with a cooling/heating rate of $3\degree$ C/min. O$_2$ flow was stopped 30 min after the furnace cooled to room temperature. The samples were taken out of the quartz tubes in air and placed into the antechamber of the glovebox as quickly as possible in order minimize contact with ambient atmosphere.
    \subsection{Synthesis of 0-Ln (Ln = Pr, Tb).}\label{sec1.3}
Polycrystalline powder samples of Li$_8$LnO$_6$ (0-Ln, Ln = Pr, Tb) were synthesized similarly to 2-Ln by intimately mixing Li$_2$O and Pr$_6$O$_{11}$ (Tb$_4$O$_7$) in molar ratio 9.6:1 (20\% excess Li$_2$O). Following the similar procedure to 2-Ln, the firing was performed at $700\degree$ C for $12$ h with a cooling/heating rate of $3\degree$ C/min.   

\clearpage

\section{Characterization.}\label{sec2}
    \subsection{Powder X-ray diffraction (PXRD).}\label{sec2.1}
\begin{figure}[h!]
    \begin{center}
        \includegraphics[width=1.\textwidth]{ 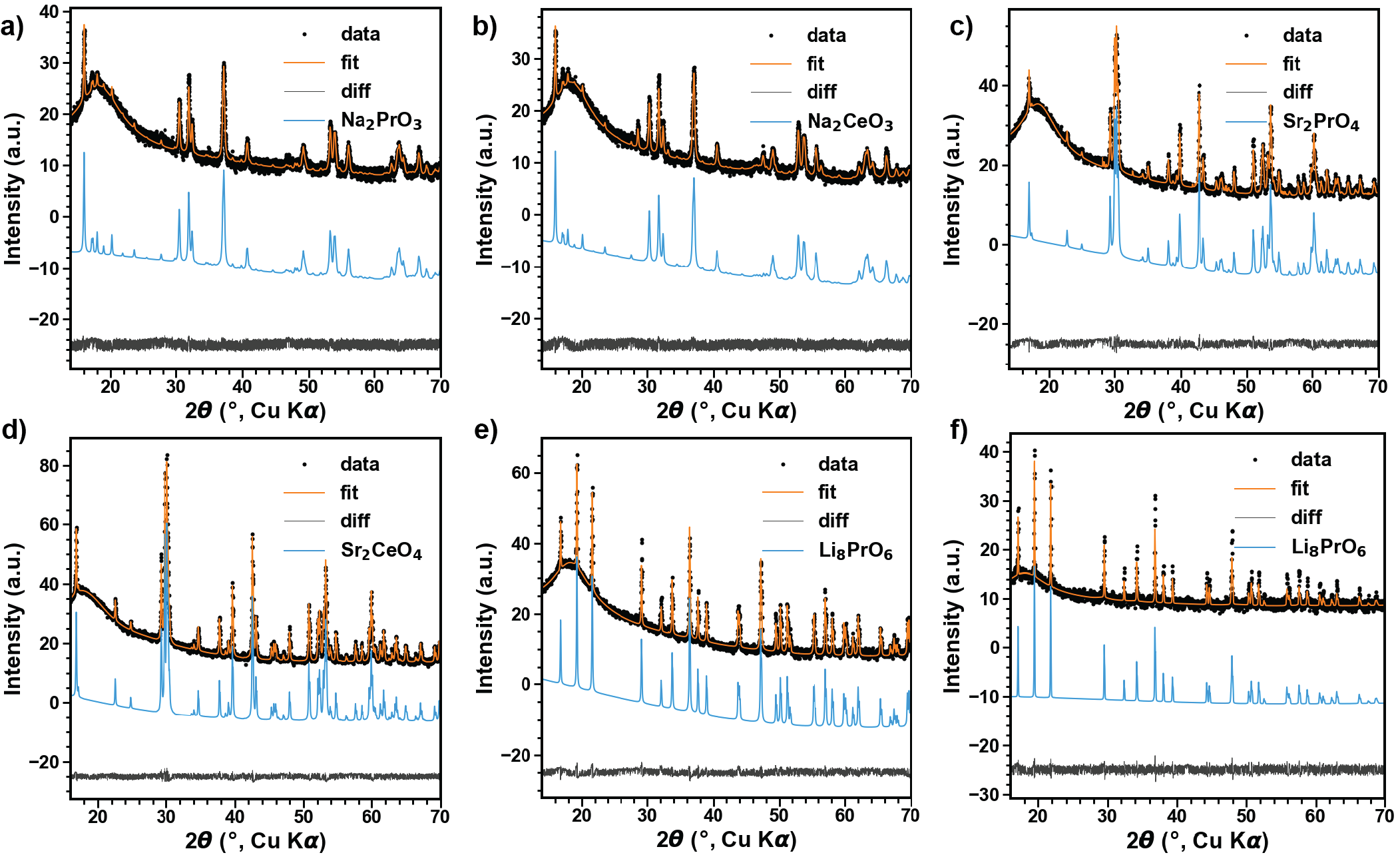}
    \end{center}
    \caption{\label{pxrd}
{\bf PXRD of different compounds.} {\bf a}, 2-Pr (Na$_2$PrO$_3$). {\bf b}, 2-Ce (Na$_2$CeO$_3$). {\bf c}, 1-Pr (Sr$_2$PrO$_4$). {\bf d}, 1-Ce (Sr$_2$CeO$_4$). {\bf e}, 0-Pr (Li$_8$PrO$_6$). {\bf f}, 0-Tb (Li$_8$TbO$_6$). Data is shown in black dots, Rietveld refinements in orange, the corresponding phases in blue, and the difference curves in grey. The broad hump near $2\theta=20 (\degree)$ corresponds to polycarbonate dome background from the sample holder. All data was collected at $T=300$ K. Quantitative Rietveld refinements to the laboratory XRD data were carried out using Bruker TOPAS 5 suite\cite{coelho2018topas}.
 	}    
\end{figure}

\begin{table}[h!]
\centering
\begin{tabular}{ccccccc}
\hline
             & 2-Pr & 2-Ce & 1-Pr & 1-Ce & 0-Pr & 0-Tb  \\ \hline

Space group  & $C_{2/c}$ & $C_{2/c}$ & $Pbam$ & $Pbam$ & $R\overline{3}$ & $R\overline{3}$\\
Point group  & $D_{2d}$ & $D_{2d}$ & $C_{2h}$ & $C_{2h}$ & $S_{6}$ & $S_{6}$\\
a(\AA) & $5.963(3)$ & $6.074(3)$ & $6.123(8)$ & $6.118(9)$ & $5.608(5)$ & $5.549(6)$\\
b(\AA) & $10.319(9)$ & $10.365(6)$ & $10.280(3)$ & $10.349(5)$ & $5.608(5)$ & $5.549(6)$\\
c(\AA) & $11.732(1)$ & $11.774(8)$ & $3.589(6)$ & $3.597(1)$ & $15.982(4)$ & $15.709(7)$\\
$\alpha(\degree)$ & $90$ & $90$ & $90$ & $90$ & $90$ & $90$\\
$\beta(\degree)$ & $109.9(1)$ & $110.1(1)$ & $90$ & $90$ & $90$ & $90$\\
$\gamma(\degree)$ & $90$ & $90$ & $90$ & $90$ & $120$ & $120$\\
$d_{Ln-Ln}$ (\AA) & $3.407(3), 3.487(6)$ & $D_{2d}$ & $3.589(6)$ & $3.597(11)$ & $5.608(5)$ & $5.549(6)$\\

\hline
\end{tabular}
\label{tab:structure}
\caption{Crystal structure information.}
\end{table}

\clearpage
    \subsection{Inelastic neutron scattering.}
Broadband inelastic neutron scattering measurements using $E_i$ = 300, 500, and 700 meV revealed a number of flat modes, common across all compounds. The flat modes are attributed to vibrational excitations and the dispersive background visible in the $E_i = 800$ meV is attributed to a hydrogen recoil line with the clear quadratic, $Q^2$, dependence typical of recoil processes and previously observed by Sensei $et al$. At energies above around 400 meV, the flat modes are in the frequency range of OH stretching mode from a OH impurity in the starting materials and identified to be $< 3$ wt\% from laboratory powder X-ray diffraction of the Na$_2$O and Li$_2$O starting materials. The OH stretches show a strong $Q$ dependence at higher $Q$ ruling them out as CEF transitions. After accounting for the OH stretching mode, we were able to identify clear crystal-electric field transitions in 0-Pr and 1-Pr. Broadband inelastic neutron scattering data was reduced and analyzed in MANTID on the SNS analysis cluster, ORNL. All diagonalization were carried out using pycrystalfield\cite{scheie2021pycrystalfield}.

\begin{figure}[h!]
    \begin{center}
        \includegraphics[width=0.7 \textwidth]{ 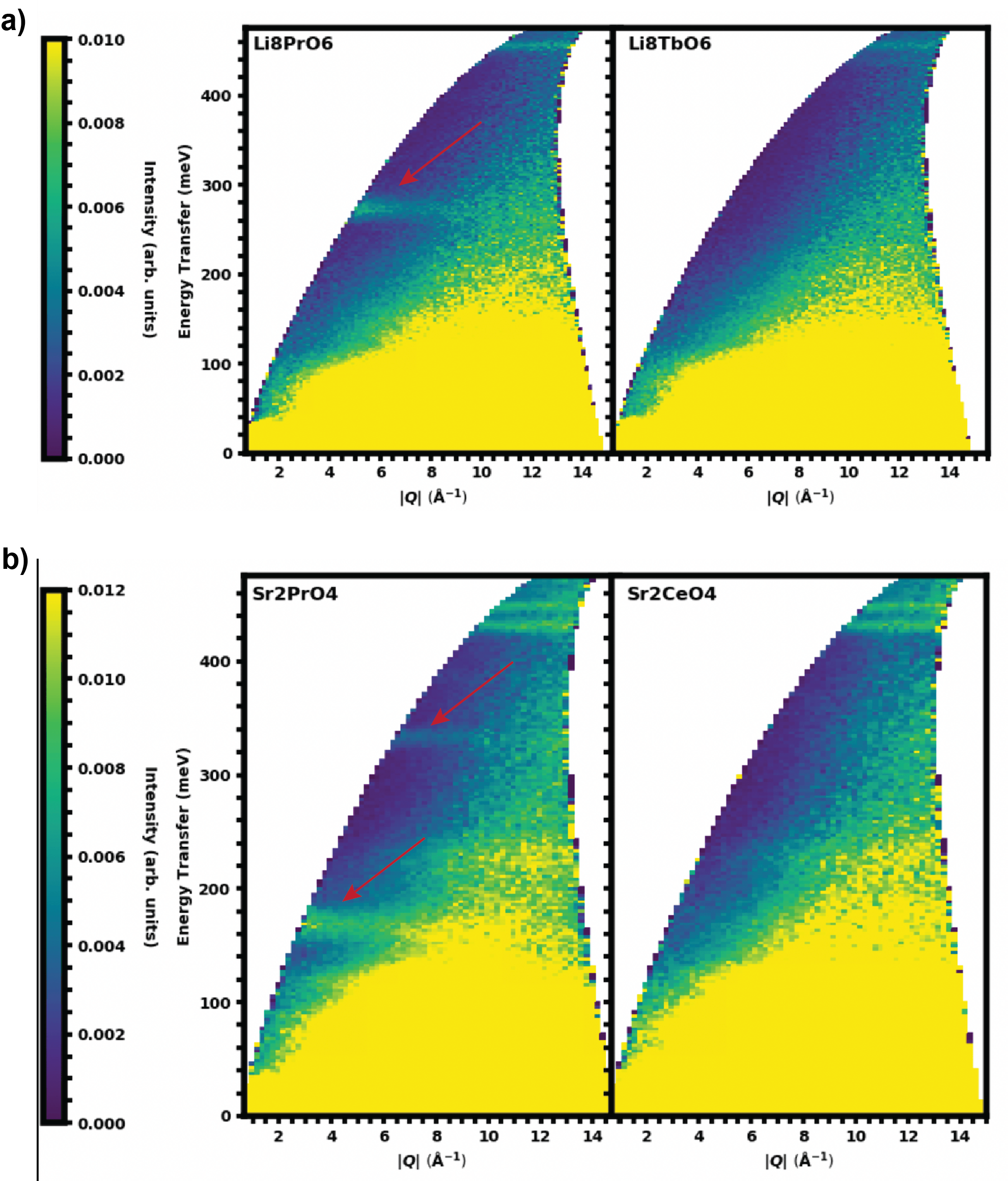}
    \end{center}
    \caption{\label{Neutron}
{\bf Overview of INS data measured on SEQUOIA.} {\bf a}, 0-Ln. {\bf b}, 1-Ln with incident energy of $E_i = 500$ meV. The CF transitions are indicated by red arrows.  
 	}    
\end{figure}

\clearpage
    \subsection{STXM O \texorpdfstring{$K$}~~edge XAS.}
The O $K$-edge STXM data were background subtracted using the MBACK algorithm in MATLAB. The data were normalized by fitting a first-order polynomial to the post-edge region of the spectrum and setting the edge jump at $541$6 eV to an intensity of 1.0. The spectra was fit to pseudo-voigt lineshapes using in-house built python scripts. Approximate peak positions were determined using first and second derivatives of the spectrum. The edge was modeled using a step function. The TEY and TFY data were normalized to the maximum of the $M_5$ edge. To facilitate comparisons to previously reported O $K$ and Pr $M_{5,4}$ edge spectra for PrO$_2$, the energy position of the step fucntion was optimized near the value used previously. 
\begin{figure}[h!]
    \begin{center}
        \includegraphics[width=1.\textwidth]{ 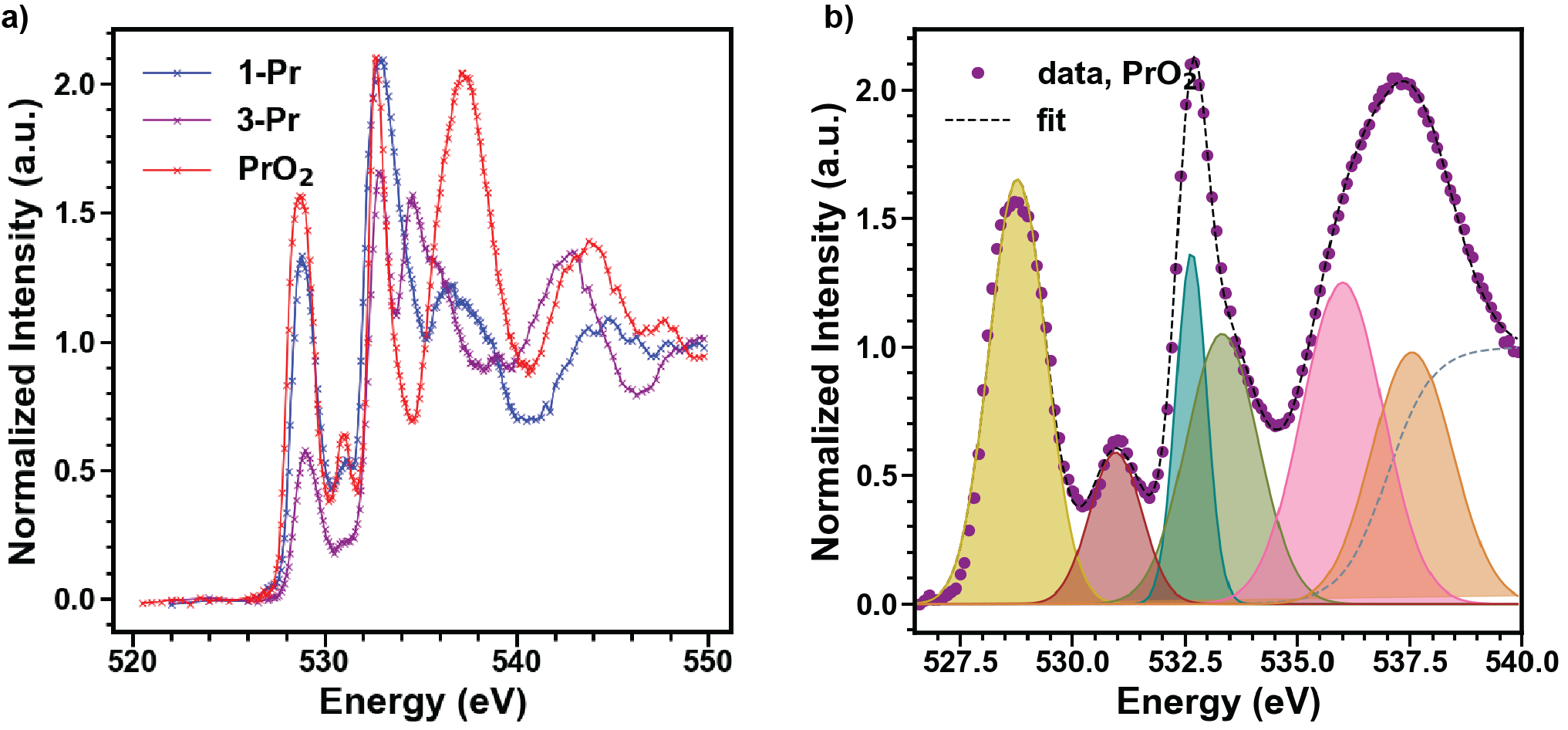}
    \end{center}
    \caption{\label{STXM K-edge}
{\bf STXM O $K$~edge XAS.} {\bf a}, STXM O $K$ edge data plotted together for 2-Pr, 0-Pr, and PrO$_2$ to show the differences in $4f$ hybridization between them. {\bf b}, Fits to PrO$_2$. Data obtained from \cite{minasian2017quantitative}. All data was collected at $T=300$ K.
 	}    
\end{figure}
    
\clearpage   
    \subsection{STXM Pr \texorpdfstring{$M$}~~edge XAS.}

\begin{figure}[h!]
    \begin{center}
        \includegraphics[width=1.\textwidth]{ 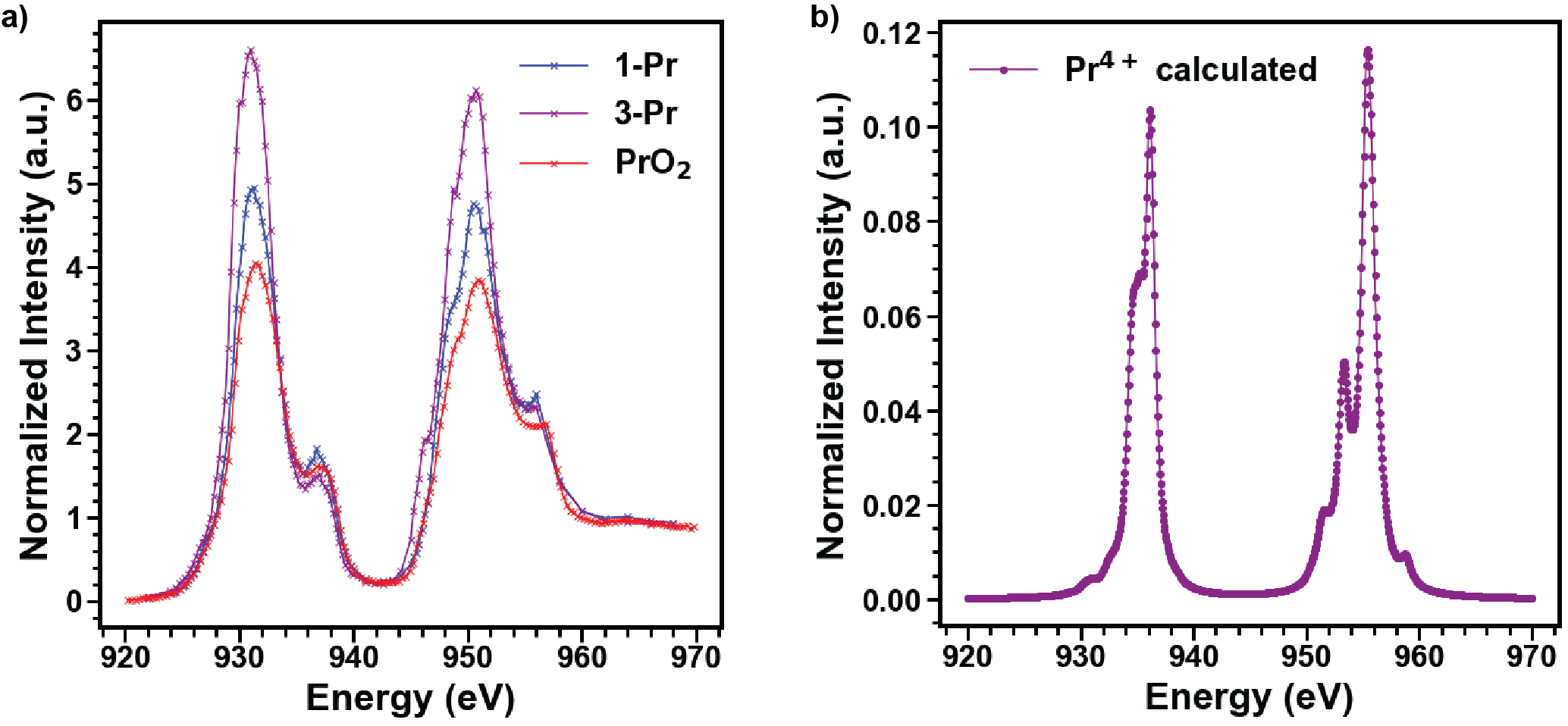}
    \end{center}
    \caption{\label{STXM M edge}
{\bf STXM Pr $M_{5,4}$~edge XAS.} {\bf a}, STXM Pr $M_{5,4}$ edge data plotted together for 2-Pr, 0-Pr, and PrO$_2$ to show the differences in peak intensities corresponding to difference in hybridization. {\bf b}, Pr $M_{5,4}$ edge XAS calculated using atomic multiplet theory\cite{cowan1981theory} for a Pr$^{4+}$ system which does not include Pr-$4f$/O-$2p$ hybridization. As explained in the main text, the calculation does not capture the satellite peaks and predicts a structured $M_4$ edge which is clearly absent in the data. PrO$_2$ data was obtained from\cite{minasian2017quantitative} . All data was collected at and calculations performed at $T=300$ K
 	}    
\end{figure}

\begin{figure}[h!]
    \begin{center}
        \includegraphics[width=1.\textwidth]{ 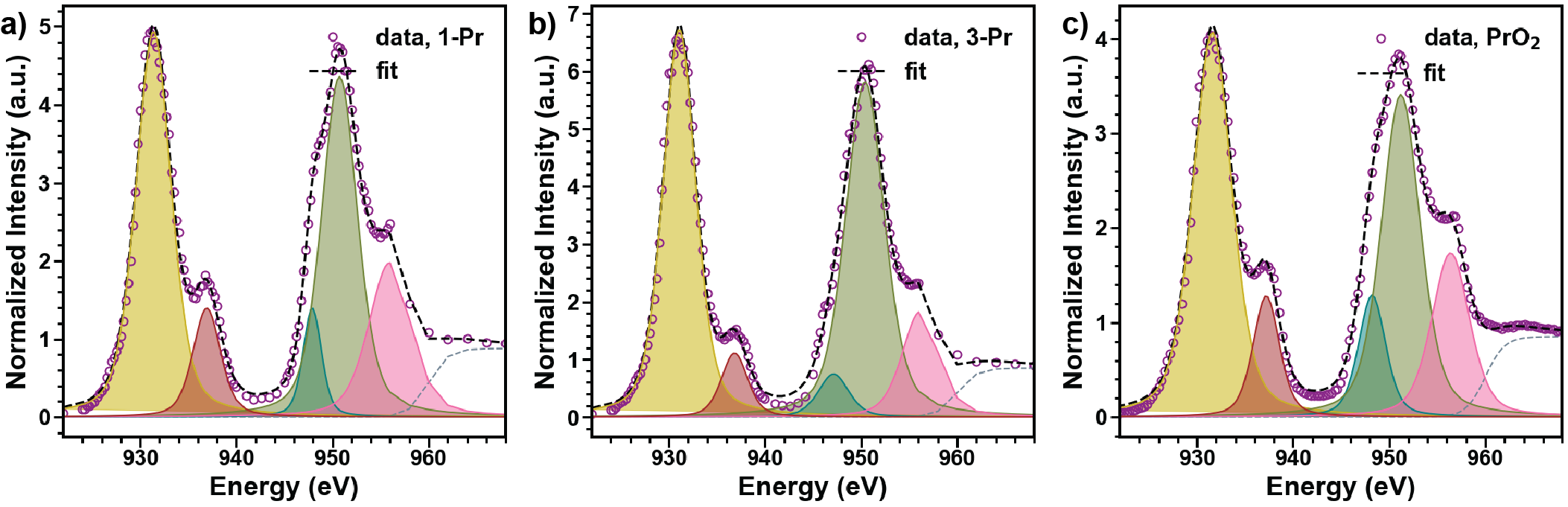}
    \end{center}
    \caption{\label{STXM M edge fits}
{\bf STXM Pr $M_{5,4}$~edge XAS fits.} {\bf a}, 2-Pr (Na$_2$PrO$_3$). {\bf b}, 0-Pr (Li$_8$PrO$_6$). {\bf c}, PrO$_2$. Fits were carried out in in-house built python scripts using pseudo-voigt functions and setting the step jump to 1.0. The data was post-edge normalized. 
 	}    
\end{figure}

\begin{table}[h!]
\centering
\renewcommand{\arraystretch}{2}
\begin{tabular}{ccccc}
\hline
             & CEF ($\Gamma_7 \rightarrow \Gamma_8$, meV)$^a$ & $I_{1s \rightarrow 4f}^b$ & $\frac{I_A}{I_{A^{\prime}}}^c$ & $\frac{I_B}{I_{B^{\prime}}}^c$ \\ \hline

PrO$_2$  & $\sim 130$ & $3.65(9)$ & $3.176(4)$ & $1.963(6)$ \\
2-Pr  & $\sim 168$ & $2.73(11)$ & $3.540(4)$ & $2.207(5)$\\
0-Pr & $\sim 260$ & $1.08(3)$ & $6.062(4)$ & $3.177(1)$\\

\hline
\end{tabular}
\label{tab:pr-pro6oh}
\flushleft{$^a$ From $ab initio$ calculations, INS, and FIRMS measurements. 
\\$^b$ From O $K$ edge XANES.
\\$^c$ From Pr $M_{5,4}$ XANES. The notations $A,B$ and $A^{\prime},B^{\prime}$ correspond to the notations in Fig3 in the main text. $I$ also corresponds to normalized intensity of the peaks and not the integrated intensity.}
\caption{Difference in degree of hybridization between PrO$_2$, 2-Pr, and 0-Pr evident from CEF transitions, O $K$ edge, and Pr $M_{5,4}$ edge XAS.}
\end{table}


\clearpage
    \subsection{Pr \texorpdfstring{$M_{5,4}$}~~edge XMCD.}
\begin{figure}[h!]
    \begin{center}
        \includegraphics[width=1.\textwidth]{ 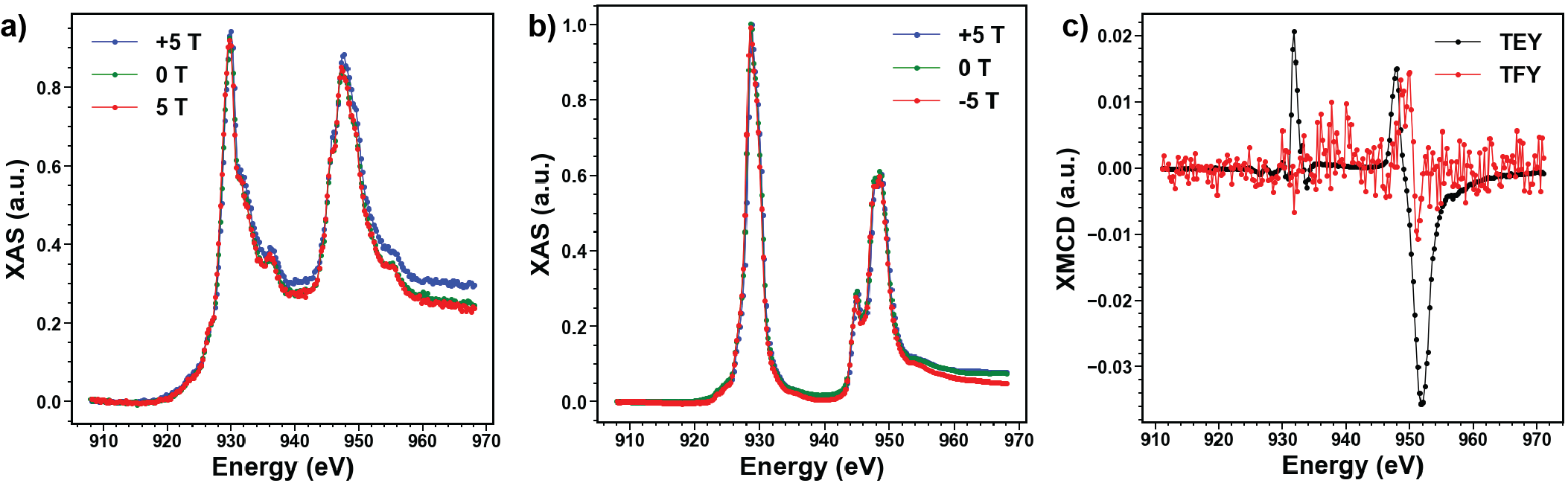}
    \end{center}
    \caption{\label{M edge 1-Pr}
{\bf Pr $M_{5,4}$~edge XMCD in 2-Pr.} {\bf a}, Total fluorescence yield (TFY) isotropic data measured under $\mu_oH=0,+5,-5$ T. As expected, the TFY data looks similar to the STXM data except for a small bump at $\sim 938$ eV which corresponds to Cu from the sample holder. {\bf b}, Total electron yield (TEY) isotropic data measured under $\mu_oH=0,+5,-5$ T. The signal from Cu is not visible in the TEY data. {\bf c}, XMCD data measured in both TEY and TFY modes. Given the poor signal of TFY, only the TEY was used for XMCD analysis in the main text. The data was normalized to the maximum of the $M_5$ edge. 
 	}    
\end{figure}    
\begin{figure}[h!]
    \begin{center}
        \includegraphics[width=1.\textwidth]{ 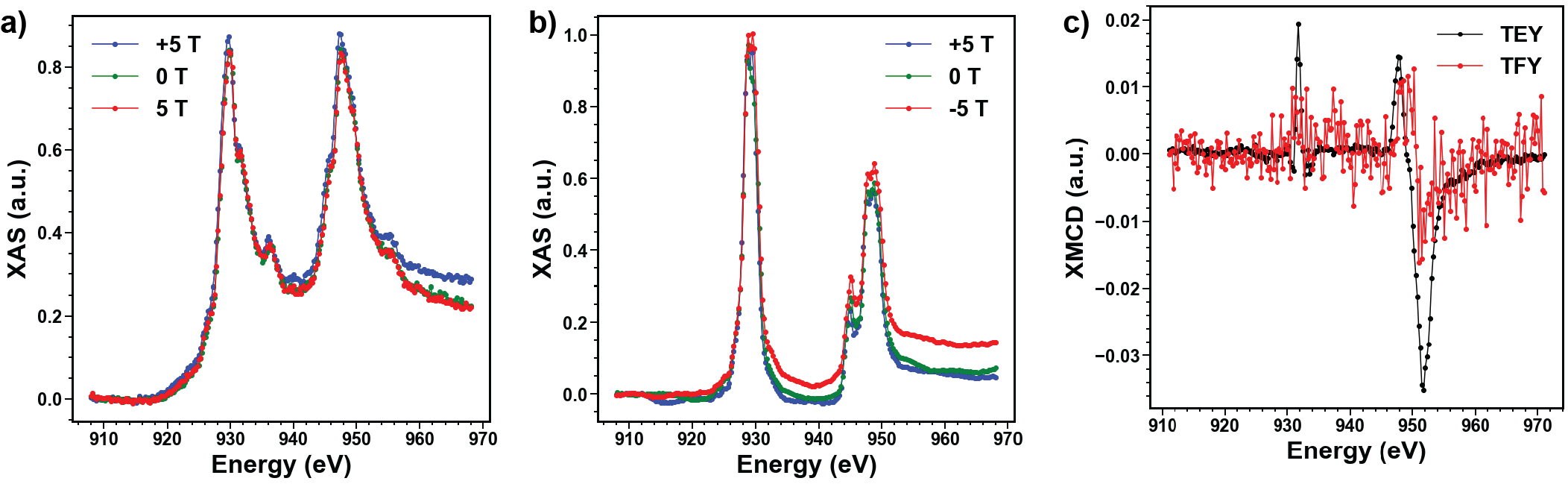}
    \end{center}
    \caption{\label{M edge 3-Pr}
{\bf Pr $M_{5,4}$~edge XMCD in 0-Pr.} {\bf a}, Total fluorescence yield (TFY) isotropic data measured under $\mu_oH=0,+5,-5$ T. As expected, the TFY data looks similar to the STXM data except for a small bump at $\sim 938$ eV which corresponds to Cu from the sample holder. {\bf b}, Total electron yield (TEY) isotropic data measured under $\mu_oH=0,+5,-5$ T. The signal from Cu is not visible in the TEY data. {\bf c}, XMCD data measured in both TEY and TFY modes. Given the poor signal of TFY, only the TEY was used for XMCD analysis in the main text. The data was normalized to the maximum of the $M_5$ edge. 
 	}    
\end{figure}

\begin{figure}[h!]
    \begin{center}
        \includegraphics[width=1.\textwidth]{ 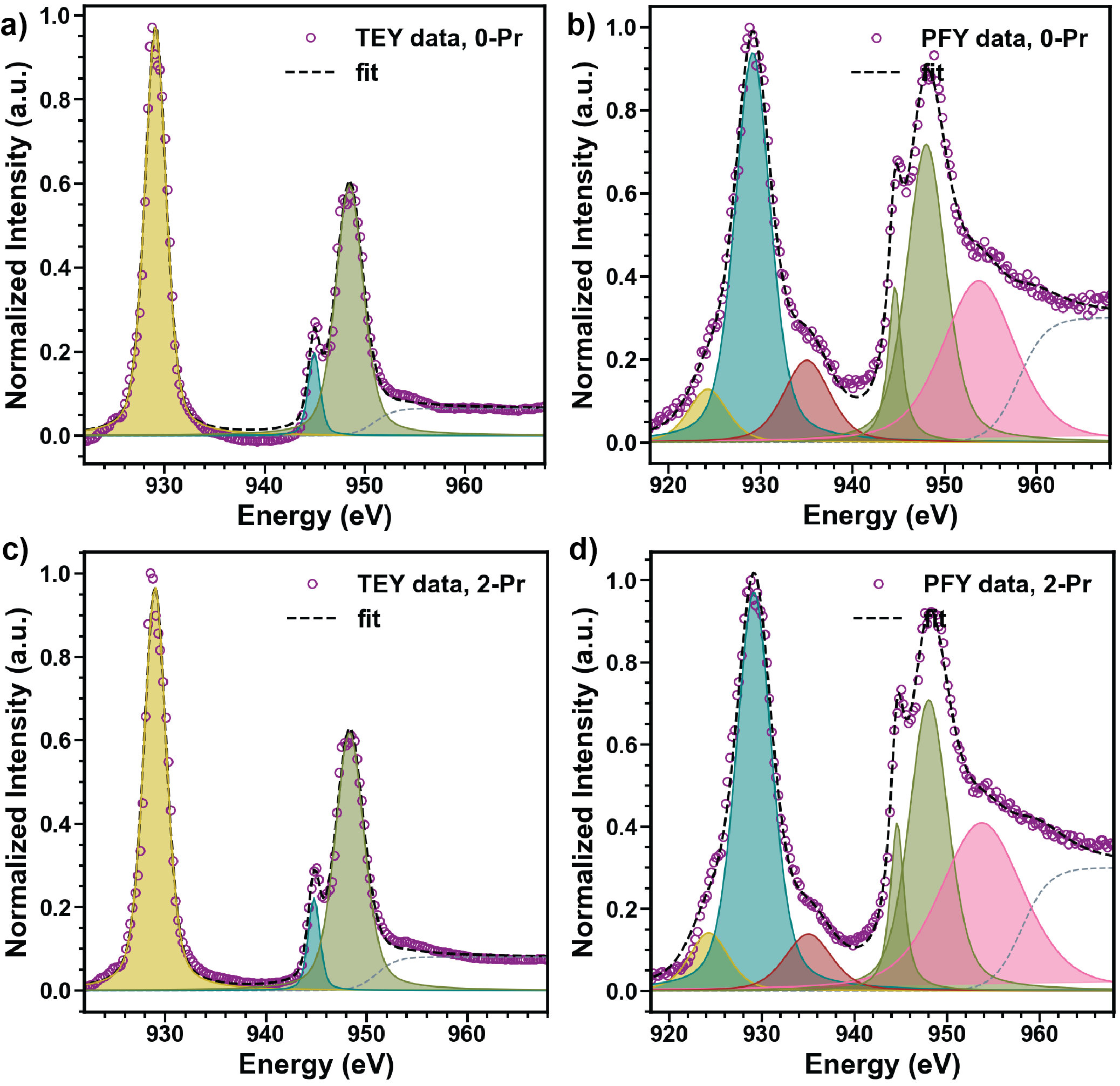}
    \end{center}
    \caption{\label{PFY}
{\bf Pr $M_{5,4}$~edge.} {\bf a}, Total electron yield (TFY) isotropic data measured under $\mu_oH=0$ T and the corresponding fits for comparisons for 0-Pr. {\bf b}, Partial fluorescence yield (PFY) isotropic data measured under $\mu_oH=0$ T and the corresponding fits for 0-Pr. As expected PFY data looks similar to the STXM data. The branching ratio (BR) for TEY is 0.52(3) and for PFY is 0.43(2) in line with STXM data. {\bf c}, Total electron yield (TFY) isotropic data measured under $\mu_oH=0$ T and the corresponding fits for comparisons for 2-Pr. {\bf d}, Partial fluorescence yield (PFY) isotropic data measured under $\mu_oH=0$ T and the corresponding fits for 2-Pr. As expected PFY data looks similar to the STXM data. The branching ratio (BR) for TEY is 0.51(4) and for PFY is 0.42(3) in line with STXM data.  
 	}    
\end{figure}
\clearpage
\section{Data Analysis.}
    \subsection{CF splitting of Pr$^{4+}$ in $|j,m_j\rangle$ basis.}
The strong spin orbit coupling of a 4$f^1$ ion, entangles the electron spin, $S = 1/2$ and high orbital angular momentum $L = 3$ to give rise to a $J = 5/2$ ground state ($^2F_{5/2}$) and a $J = 7/2$ excited state ($^2F_{7/2}$). The sixfold degeneracy of the $^2F_{5/2}$ ground state is removed under the crystal field. Under a highly symmetric $O_h$ symmetry, the $^2F_{5/2}$ ground state is split into a doublet $\Gamma_7$ and a quartet $\Gamma_8$. Any deviation from the $O_h$ symmetry will remove the degeneracy of the $\Gamma_8$ state resulting a maximum splitting of $^2F_{5/2}$ state into three Kramers doublets. The Kramers doublets are given by $\Gamma_7^{\pm} = \sin\theta|\frac{5}{2},\pm\frac{5}{2}\rangle + \cos\theta|\frac{5}{2},\mp\frac{3}{2}\rangle$, where $\sin^2\theta\sim1/6$, $\Gamma_{8,1}^{\pm} = |\frac{5}{2},\pm\frac{1}{2}\rangle$, and $\Gamma_{8,2}^{\pm} = \alpha|\frac{5}{2},\pm\frac{3}{2}\rangle + \sqrt{1-\alpha^2}|\frac{5}{2},\mp\frac{5}{2}\rangle$, respectively. The CF Hamiltonian for a perfect $O_h$ symmetry is written as $\mathcal{\hat{H}}_{\rm CEF} = B^0_4\hat{O}^0_4+ B^4_4\hat{O}^4_4 + B^0_6\hat{O}^0_6 + B^4_6\hat{O}^4_6$ where $B^m_n$ are the second, fourth, and sixth order terms and $\hat{O}^n_m$ are the corresponding Stevens operator equivalents\cite{stevens1952matrix}. Further constraints includes, $B^4_4 = 5*B^0_4$ and $B^4_6 = -21*B^4_6$.In the $J_{\rm eff}\!=\!1/2$ limit, the maximum allowed terms in the CF Hamiltonian are less than 2$J$ meaning that the sixth order terms is zero; $B^6_n$ = 0, evaluated in the total angular momentum $|j,m_j\rangle$ basis. In this limit, the essential physics is limited to $^2F_{5/2}$ as is the case for Ce$^{3+}$ systems. 
    \subsection{CF splitting of Pr$^{4+}$ in $|m_l,m_s\rangle$ basis.}
The $O_h$ crystal field splits the seven $f$ orbitals to ground state $a_{2u}$, and excited triply degenerate $t_{1u}$ and $t_{2u}$ states. In the presence of spin-orbit coupling, the seven $f$ orbitals mix yielding 14 KD in line with the 14 states extracted from $|j,m_j\rangle$ states. In the $|m_l,m_s\rangle$ basis, the nature of the $\Gamma_7$ KD is given as $|\Gamma_7^{\pm}\rangle=A~\sqrt{\frac{6}{7}}|\mp3,\pm\frac{1}{2}\rangle-~B~\sqrt{\frac{1}{7}}|\mp2,\mp\frac{1}{2}\rangle~+~C~\sqrt{\frac{2}{7}}|\pm1,\pm\frac{1}{2}\rangle-~D~\sqrt{\frac{5}{7}}|\pm2,\mp\frac{1}{2}\rangle$, where $\alpha=\frac{A^2+B^2}{C^2+D^2}\sim0.18$ and $(A^2/B^2)^{\rm \Gamma_7}\sim2.6$ and $(C^2/D^2)^{\rm \Gamma_7}\sim0.33$. The first two components of $\Gamma_7$ KD ($m_l=-3,-2$) identifies with being derived from $|\frac{5}{2},\pm\frac{5}{2}\rangle, |\frac{7}{2},\pm\frac{5}{2}\rangle$, states in $|j,m_j\rangle$ basis, while the last components ($m_l=1,2$) identifies with being derived from $|\frac{5}{2},\pm\frac{3}{2}\rangle, |\frac{7}{2},\pm\frac{3}{2}\rangle$ states. In the $J_{\rm eff}\!=\!1/2$ limit, the contributions from $|\frac{7}{2},\pm\frac{5}{2}\rangle, |\frac{7}{2},\pm\frac{5}{2}\rangle$ states are negligible with the essential physics limited to $^2F_{5/2}$ SOC manifold as described above. In the $|m_l,m_s\rangle$ basis framework the $O_h$ CF Hamiltonian must be diagonalized with the entire set of 14 $LS$ basis states and thereby making higher-order terms as non-zero, $B^6_n\neq0$, while the constraints for $B^4_4$ and $B^4_6$ still apply in the $O_h$ symmetry. This framework overcomes the point-charge model established in the $|j,m_j\rangle$ basis and gives a better approximation of the ground state for covalent lanthanide systems. The parameter $\alpha$ defines the ratio of $|j,\pm\frac{5}{2}\rangle$ to $|j,\pm\frac{3}{2}\rangle$. For the original $\Gamma_7$ KD, $\alpha\sim0.25$ indicates the ground state wavefunction is primarily defined by the $m_l=1,2$ ($|j,\pm\frac{3}{2}\rangle$) components which agrees well with the $\Gamma_7$ wavefucntion derived in the $|j,m_j\rangle$ basis above. 
    \subsection{Intermediate coupling for Pr$^{4+}$ in $|m_l,m_s\rangle$ basis.}
\begin{figure}[h!]
    \begin{center}
        \includegraphics[width=1.\textwidth]{ 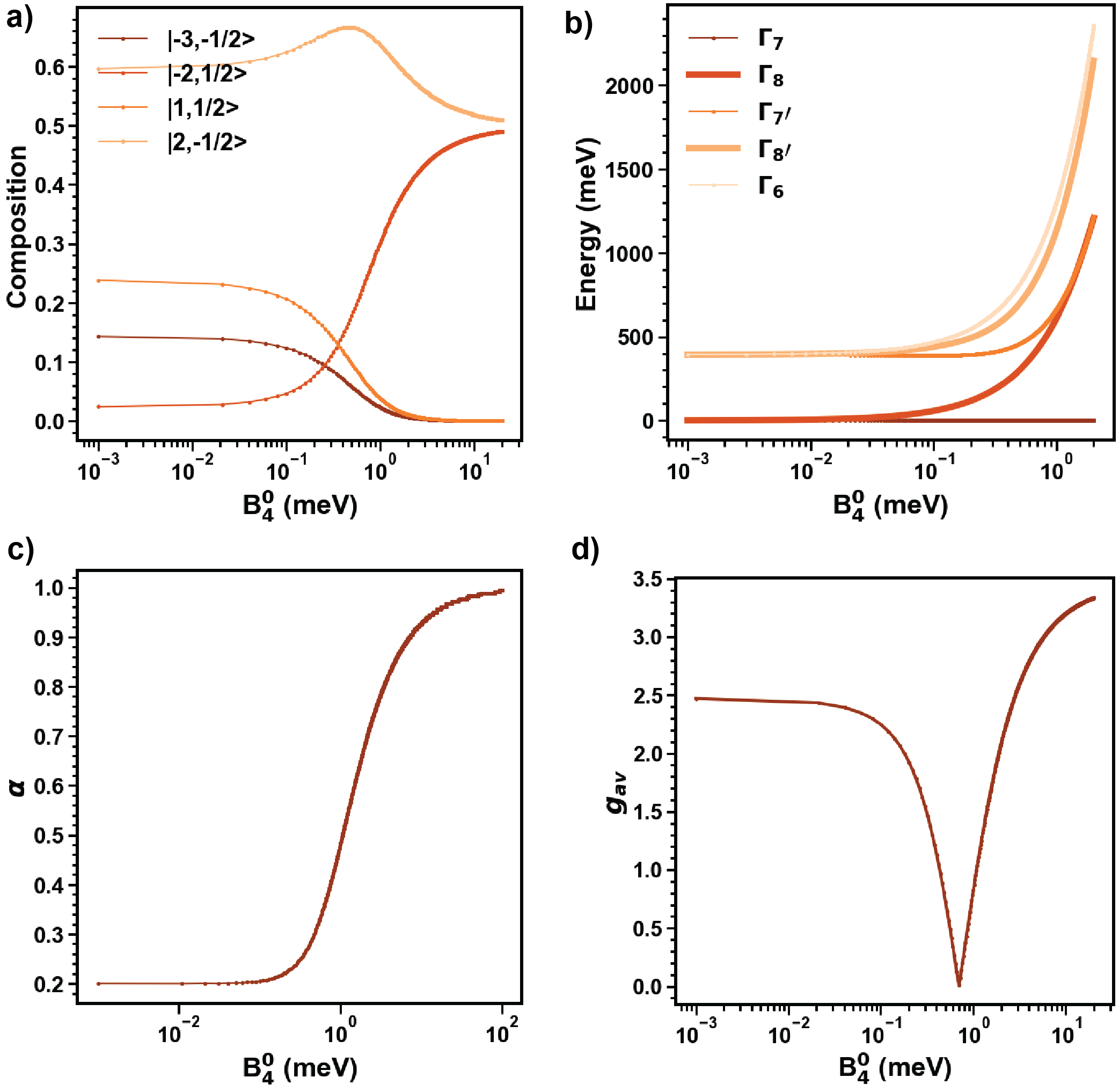}
    \end{center}
    \caption{\label{IC}
{\bf Evolution of single-ion properties as a function of $B^0_4$ with $B^0_6 = 0$.} {\bf a}, Compsition of the ground state wavefucntion. {\bf b}, Eigen energies of the different eigen states relative to the ground state set to 0 meV. The thickness of the lines correspond to the degeneracy of the eigen states. {\bf c}, Evolution of $\alpha$. {\bf d}, Evolution of $g_{av}$. 
 	}    
\end{figure}

\begin{figure}[h!]
    \begin{center}
        \includegraphics[width=1.\textwidth]{ 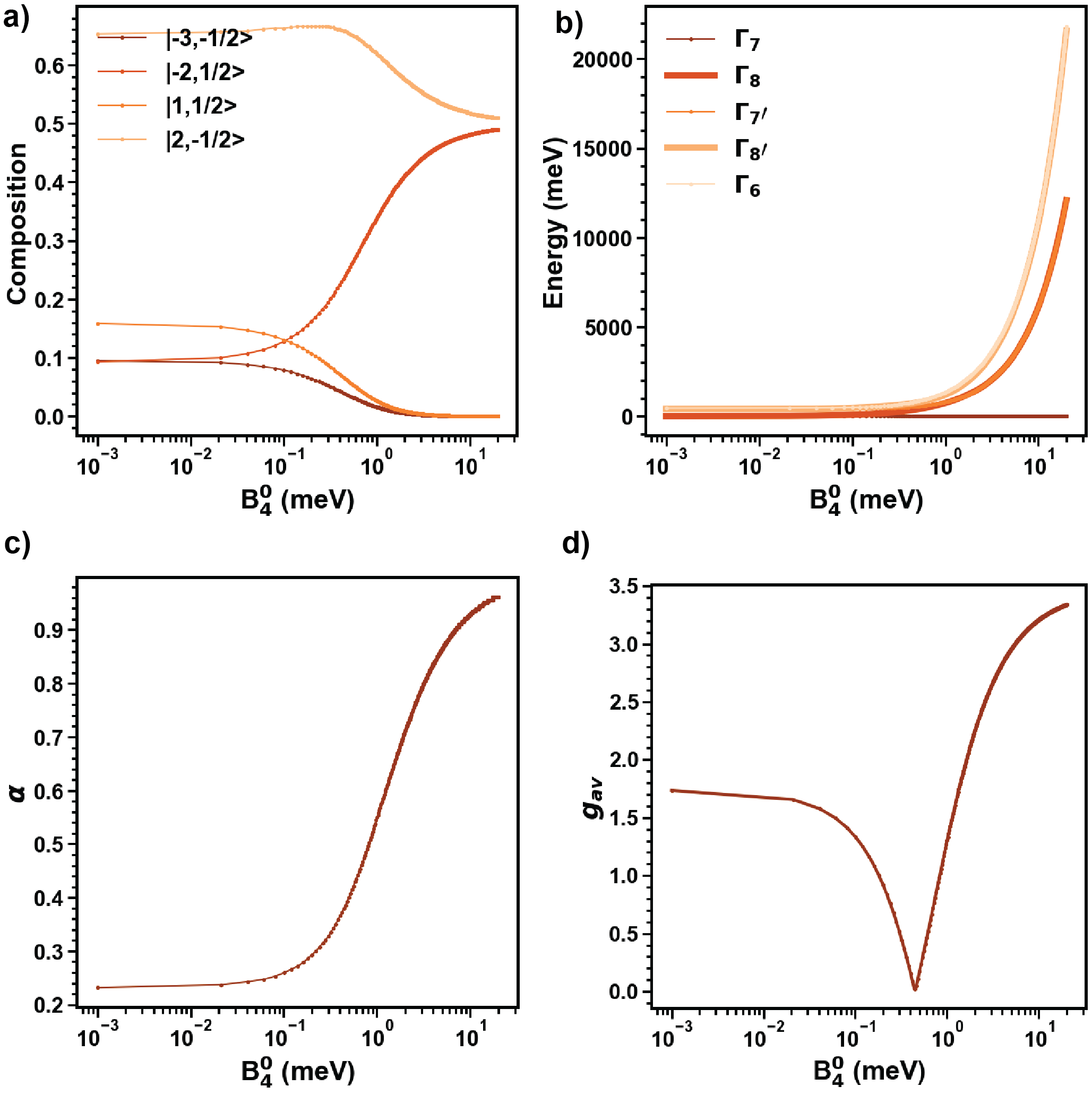}
    \end{center}
    \caption{\label{IC_0p01}
{\bf Evolution of single-ion properties as a function of $B^0_4$ with $B^0_6 = 0.01$.} {\bf a}, Compsition of the ground state wavefucntion. {\bf b}, Eigen energies of the different eigen states relative to the ground state set to 0 meV. The thickness of the lines correspond to the degeneracy of the eigen states. {\bf c}, Evolution of $\alpha$. {\bf d}, Evolution of $g_{av}$. 
 	}    
\end{figure}

\begin{figure}[h!]
    \begin{center}
        \includegraphics[width=1.\textwidth]{ 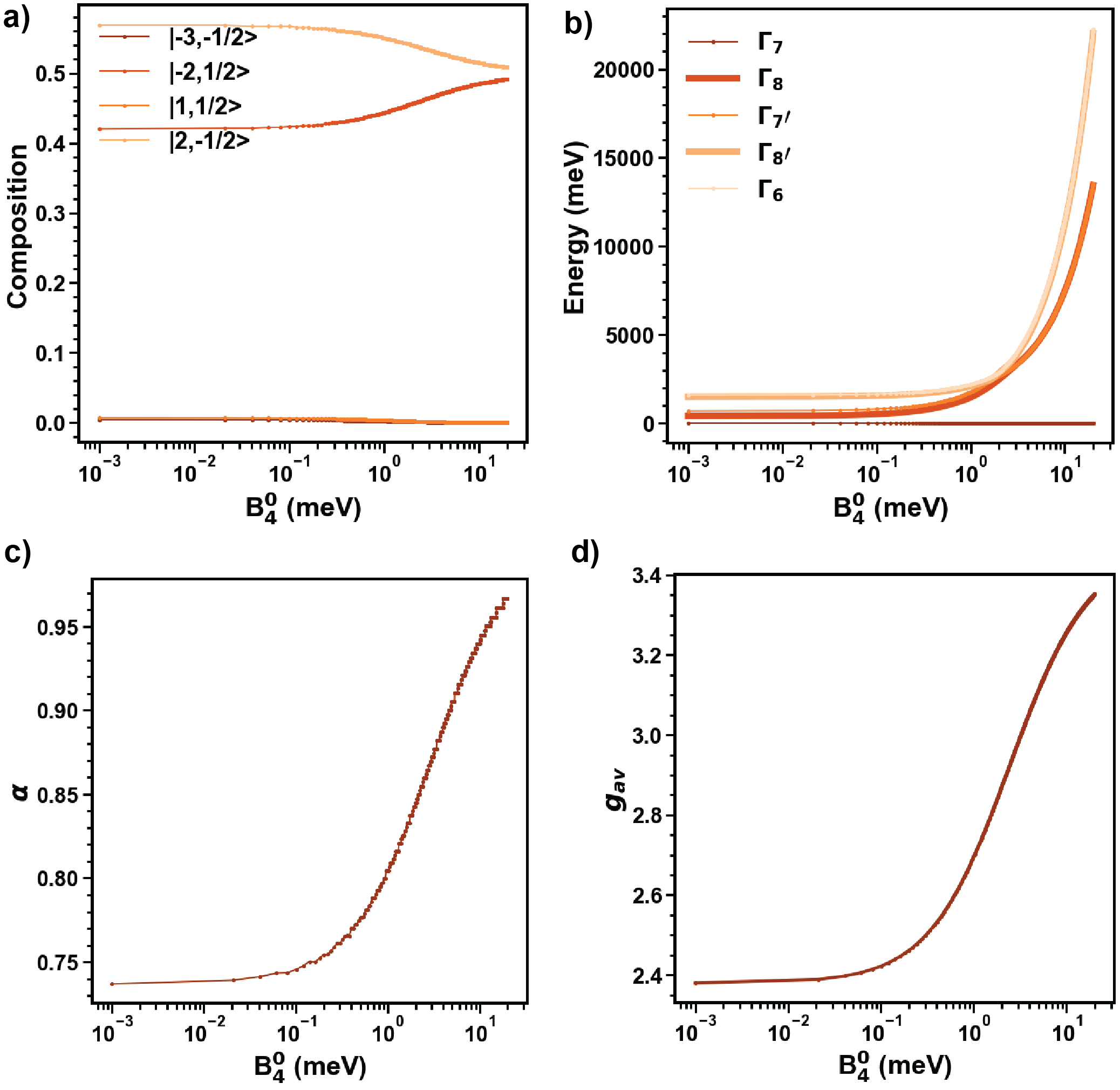}
    \end{center}
    \caption{\label{IC_0p1}
{\bf Evolution of single-ion properties as a function of $B^0_4$ with $B^0_6 = 0.1$.} {\bf a}, Compsition of the ground state wavefucntion. {\bf b}, Eigen energies of the different eigen states relative to the ground state set to 0 meV. The thickness of the lines correspond to the degeneracy of the eigen states. {\bf c}, Evolution of $\alpha$. {\bf d}, Evolution of $g_{av}$. 
 	}    
\end{figure}

Given the large CF energy scale for Pr$^{4+}$ as described in the main text, CF and SOC interactions now compete with each other. Therefore, CF cannot be considered as a perturbation on the SOC energy scale and thereby making the $J_{eff}\!=\!1/2$ picture invalid. The presence of competing interactions drastically changes the single-ion picture which further affects the macroscopic properties of the system as described in the main text. In order to understand the implications of intermediate coupling, we study the evolution of the single-ion properties of Pr$^{4+}$ in the toy model Hamiltonian $\mathcal{\hat{H}}_{\rm CEF}^{Pr} = B^0_4\hat{O}^0_4 + B^4_4\hat{O}^4_4 + B^0_6\hat{O}^0_6 + B^4_6\hat{O}^4_6$ constrained by $O_h$ symmetry for an isolated PrO$_6$ unit. We calculate, the eigen energies, ground state wavefucntion composition, $\alpha$, and $g_{av}$ as a function of $B^0_4$ for fixed values of $B^0_6$ as shown in figs S\ref{IC},S\ref{IC_0p01}, and S\ref{IC_0p1}. Here, $B^0_4 > 0$ as is expected for a six coordinate system. $B^0_4 < 0$ flips the first excited state quartet as the ground state and the $\Gamma_7$ KD as the excited state and is expected for a eight coordinate system with cubic symmetry as is the case for PrO$_2$. With $B^0_6$ = 0 (Fig. S\ref{IC}), for small values of $B^0_4$, the system can be considered as a traditional lanthanide where the $J_{\rm eff}\!=\!1/2$ limit still applies. As expected the value of $\alpha$ remains close to 0.25 with the eigen states split as expected for the $O_h$ CF. As $B^0_4$ increases, the system begins to deviate from the $J_{\rm eff}\!=\!1/2$ limit and moves towards the $S_{\rm eff}\!=\!1/2$ where the $\Delta_{CF} >> \zeta_{SOC}$ as described in the main text. For nonphysically large values of $B^0_4$, the eigen states relax to 3 states where the ground state KD corresponds $a_{2u}$, the first excited state with three degenerate KD corresponds to triply degenerate $t_{1u}$ and the second excited state with three degenerate KD corresponds to triply degenerate $t_{2u}$. The values of $\alpha$ tends towards 1 as we move towards the $S_{\rm eff}\!=\!1/2$ limit. This indicates that, with increase in CF energy scale, the $|j,\pm\frac{5}{2}\rangle$ character adds on to the ground state $\Gamma_7$ KD with the final wavefunction in the $S_{\rm eff}\!=\!1/2$ limit having equal contribution from $|j,\pm\frac{5}{2}\rangle$ and $|j,\pm\frac{3}{2}\rangle$ states. This is also evident from the evolution of $g_{av}$ which shows a dip to almost zero and increases again. As described in the main text, Pr$^{4+}$ systems do exhibit unusually small $g$ values which is in line with intermediate coupling scheme established here.

    \subsection{CF Hamiltonian for Pr$^{4+}$ in 1-Pr.}
We begin constraining the Hamiltonian for 1-Pr under a perfect $O_h$ CF while the real symmetry is $C_{2h}$ in order to reduce the number of parameters for fitting, $\mathcal{\hat{H}}_{\rm CEF}^{1-Pr} = B^0_4\hat{O}^0_4 + B^4_4\hat{O}^4_4 + B^0_6\hat{O}^0_6 + B^4_6\hat{O}^4_6$, where $B^4_4$, and $B^4_6$ were constrained under $O_h$ symmetry. Also, $B^0_4$ was constrained to be $> 0$ as described in the above section for a six coordinate system. The energy scale of the uncommonly large $\Delta_{CEF}$ in 1-Pr is comparable in magnitude to $\zeta_{SOC}$ of Pr$^{4+}$. In the $J_{\rm eff}\!=\!1/2$ limit, treating the Pr$^{4+}$ ion as isoelectronic Ce$^{3+}$, $\mathcal{\hat{H}}_{CEF}^{1-Pr}$ was fit to the observed three transitions in INS data. However, we find that these calculations fail to accurately describe the experimental thermo-magnetic data often overestimating. Therefore, we move to intermediate coupling where we diagonalize the  $\mathcal{\hat{H}}_{\rm CEF}^{1-Pr}$ with the entire set of 14$LS$ basis states as described above. Initial guesses for the steven's coefficients $B^0_4$ and $B^0_6$ were obtained in the $|j,mj\rangle$ basis by setting the first excited state to $E_1^{\rm 1-Pr} = 168$ meV. We note here that, point change based estimation of Steven's coefficients is not appropriate for Pr$^{4+}$ given the anomalously large Pr-$4f$/O-$2p$ covalency. With the initial guesses for $B^0_4$ and $B^0_6$, we start fitting the susceptibility data ($T>50$ K, to avoid the region with short-range correlations) and eigen energies and degeneracies to the Hamiltonian $\mathcal{\hat{H}}_{\rm CEF}^{1-Pr}$. With the newly estimated values for $B^0_4$ and $B^0_6$, we begin to relieve the cubic constraints on $B^4_4$ and $B^4_6$ to account for the slight distortion from perfect $O_h$ symmetry. Again, fitting to the susceptibility and  eigen energies and degeneracies yields newly estimated values for the stevens coefficients. However, to account for the full distortion from the $O_h$ symmetry, we introduce $B^0_2$ parameter resulting in a total of 5 independent variables to be fit with the final Hamiltonian being $\mathcal{\hat{H}}_{\rm CF}^{1-Pr} = B^0_2 \hat{O}^0_2 + B^0_4\hat{O}^0_4+ B^4_4\hat{O}^4_4 + B^0_6\hat{O}^0_6 + B^4_6\hat{O}^4_6$. Although, the true symmetry of 1-Pr ($C_{2h}$) requires $|m| = 2,6$ in addition to $|m| = 0,4$ (in $B_n^m$ coefficients), any mixing induced by these parameters would not induce any further loss of degeneracy and hence their effects can be parameterized with $|m| = 0,4$ parameters. Therefore, we use the truncated Hamiltonian $\mathcal{\hat{H}}_{\rm CF}^{1-Pr}$. The final fitting was carried out by providing different weights to susceptibility and eigen energies. The final fit parameters and results are provided in Table S\ref{tab:fit}. This yields a set of new KD's with the ground state wavefucntion expressed as a "renormalized" $\Gamma_7$ with $\alpha^{1-Pr} = 0.36$. The ground state wavefucniton is given as $|\Gamma_7^{\pm}\rangle=0.428~|\mp3,\pm\frac{1}{2}\rangle-~-0.293~|\mp2,\mp\frac{1}{2}\rangle~+~-0.344~|\pm1,\pm\frac{1}{2}\rangle-~0.783~|\pm2,\mp\frac{1}{2}\rangle$.This yields a slightly easy-plane anisotropic $g$ with $g_{xy}^{1-Pr} = 1.37$ and $g_z^{1-Pr} = 0.79$. 
    \subsection{CF Hamiltonian for Pr$^{4+}$ in 0-Pr.}
We then constrained the Hamiltonian for 0-Pr using a similar method established for 1-Pr. The Hamiltonian was written as $\mathcal{\hat{H}}_{\rm CF}^{0-Pr} = B^0_2 \hat{O}^0_2 + B^0_4\hat{O}^0_4+ B^4_4\hat{O}^4_4 + B^0_6\hat{O}^0_6 + B^4_6\hat{O}^4_6$. Again, the removing the cubic constraints and introduction of $B^0_2$ parameters is essential to account for the distortion from perfect $O_h$ symmetry. The final fit parameters and results are provided in Table S\ref{tab:fit}. This yields a set of new KD's with the ground state wavefucntion expressed as a "renormalized" $\Gamma_7$ with $\alpha^{0-Pr} = 0.22$. The ground state wavefucniton is given as $|\Gamma_7^{\pm}\rangle=-0.278~|\mp3,\pm\frac{1}{2}\rangle-~0.314~|\mp2,\mp\frac{1}{2}\rangle~+~0.391~|\pm1,\pm\frac{1}{2}\rangle-~-0.819~|\pm2,\mp\frac{1}{2}\rangle$. This yields a slightly easy-axis anisotropic $g$ with $g_{xy}^{1-Pr} = 0.63$ and $g_z^{1-Pr} = 1.1$ and is comparable to $g$ values extracted from first principles calculations($g_{xy} \approx 0.7$, $g_z \approx 1.3$). 
    \subsection{CF Hamiltonian for Pr$^{4+}$ in 2-Pr.}
We then constrained the Hamiltonian for 2-Pr using a similar method established for 1-Pr. The Hamiltonian was written as $\mathcal{\hat{H}}_{\rm CF}^{0-Pr} = B^0_2 \hat{O}^0_2 + B^0_4\hat{O}^0_4+ B^4_4\hat{O}^4_4 + B^0_6\hat{O}^0_6 + B^4_6\hat{O}^4_6$. Again, the removing the cubic constraints and introduction of $B^0_2$ parameters is essential to account for the distortion from perfect $O_h$ symmetry. The final fit parameters (fitting to susceptibility for $T > 50$ K) and results are provided in Table S\ref{tab:fit}. This yields a set of new KD's with the ground state wavefucntion expressed as a "renormalized" $\Gamma_7$ with $\alpha^{2-Pr} = 0.38$. The ground state wavefucniton is given as $|\Gamma_7^{\pm}\rangle=-0.407~|\mp3,\pm\frac{1}{2}\rangle-~0.331~|\mp2,\mp\frac{1}{2}\rangle~+~0.351~|\pm1,\pm\frac{1}{2}\rangle-~-0.776~|\pm2,\mp\frac{1}{2}\rangle$. This yields a slightly easy-axis anisotropic $g$ with $g_{xy}^{1-Pr} = 1.25$ and $g_z^{1-Pr} = 0.37$ and is comparable to $g$ values extracted from first principles calculations($g_{xy} \approx 1.4$, $g_z \approx 0.4$).
    \subsection{XMCD sum rule analysis.}
The quantitative sum rule analysis proposed for XAS and XMCD measurement relates the expectation values of the spin ($\langle S_z \rangle$), orbital $\langle L_z \rangle$), and dipole term ($\langle T_z \rangle$) in the valence state. The expression developed by Thole and Carra are given as\cite{thole19853d,tripathi2018xmcd,thole1992x,carra1993x}:
\begin{align}
    \frac{\langle L_z \rangle}{3n_h} = -\frac{\int_{M_4+M_5}\Delta\mu(E)dE}{\frac{3}{2}\int_{M_4+M_5}\mu(E)dE}
\end{align}

\begin{align}
    \langle S_{eff} \rangle = -\frac{3}{2}n_h\frac{\int_{M_5}\Delta\mu(E)dE - \frac{3}{2}\int_{M_4}\Delta\mu(E)dE}{\frac{3}{2}\int_{M_4+M_5}\mu(E)dE}
\end{align}

where, $\langle S_{eff} \rangle =$ $\langle S_z \rangle$ $+ 3 \langle T_z \rangle$, $\mu(E)$ is the energy dependence of the isotropic XAS, $\Delta \mu(E)$ is the energy dependence of the XMCD, and $n_h$ is the number of holes in the system.Thus once can estimate the orbital and spin moments as $\mu_{orbital} = -~\langle L_z \rangle~\mu_B$,  $\mu_{spin} = -2$~$\langle S_z \rangle~\mu_B$, and $\mu_{total} = \mu_{orbital} + \mu_{spin}$. Evaluation of $\langle L_z \rangle$ is straightforward yielding values reported in the main text. However, evaluation of $\langle S_z \rangle$ requires quantitative information about $\langle T_z \rangle$, In most cases, the dipole term is negligible, however for lanthanides with unquenched orbital angular momentum, the dipole term is significant. Therefore, we use the macroscopic bulk magnetization measured at $\mu_{oH} = 5$ T and $T=20$ K to  extract the absolute total moment ($\mu_{total}$). From these values, we can extract $\mu_{spin}$ based on the relation $\mu_{total} = \mu_{orbital} + \mu_{spin}$. With this, we can estimate the magnetic dipole contribution based on $\langle S_{eff} \rangle =$ $\langle S_z \rangle$ $+ 3 \langle T_z \rangle$ with out any sophisticated theoretical modeling.

\clearpage
\clearpage
\clearpage

\begin{table}[h!]
\centering
\renewcommand{\arraystretch}{1}
\begin{tabular}{ccccc}
\hline
             &$\Gamma_7$ & 0-Pr & 1-Pr & 2-Pr \\ \hline

$B^0_2$  & * & $1.3(3)$ & $-12.43$ & $-5.26$  \\
$B^0_4$  & * & $0.54(1)$ & $0.76$ & $0.38$ \\
$B^4_4$ & * & $2.29(1)$ & $3.43$ & $1.85$ \\
$B^0_6$ & * & $-0.007(1)$ & $-0.001$ & $0.003$ \\
$B^4_6$ & * & $0.11(1)$ & $0.45$ & $0.112$ \\
$\zeta_{SOC}$ & * & $112$ & $112$ & $112$ \\
$g_{av}$ & $\approx1.4$ & $\approx 0.9$ & $\approx 1.1$ & $\approx 1.1$ \\
A$^a$ & $0.352$ & $0.241$ & $0.428$ & $0.407$ \\
B$^a$ & $0.215$ & $0.331$ & $0.293$ & $0.331$ \\
C$^a$ & $0.454$ & $0.363$ & $0.344$ & $0.351$ \\
D$^a$ & $0.79$ & $0.837$ & $0.783$ & $0.776$ \\
$\alpha$$^b$ & $0.18$ & $0.22$ & $0.36$ & $0.37$ \\
A$^2$/B$^2$ & $2.6$ & $0.53$ & $2.13$ & $1.51$ \\
C$^2$/D$^2$ & $0.33$ & $0.18$ & $0.19$ & $0.20$ \\
\hline
\end{tabular}
\flushleft{$^a$ Coefficients for the ground state wavefucntions as defined in Fig. 1 caption of the main text 
\\$^b$ $\alpha$ as defined above}
\caption{Fit parameters for the three different materials. The small value of $B^0_2$ for 0-Pr is indicative of how close the PrO$O_6$ unit is close to a perfect $O_h$ symmetry. The true symmetry being $S_6$ where only a mirror symmetry is broken from the ideal $O_h$ symmetry. Furthermore, the relatively large value of $B^0_6$ for 0-Pr is indicative of the first excited state being almost 1.5 larger ($\approx266$ meV) than for 1-Pr and 2-Pr ($\approx168$ meV). All units are in meV.}
\label{tab:fit}
\end{table}


\begin{table}[h!]
\centering
\renewcommand{\arraystretch}{1}
\begin{tabular}{ccccccc}
\hline
             & \multicolumn{2}{c}{0-Pr} & \multicolumn{2}{c}{1-Pr} & \multicolumn{2}{c}{2-Pr} \\ \hline
             & $E^m$ & $E^c$ & $E^m$ & $E^c$ & $E^m$ & $E^c$ \\ \hline

$KD 1$  & $0$ & $0$ & $0$ & $0$ & $0$ & $0$  \\
$KD 2$  & $267$ & $255.1$ & $168$ & $160.8$ & $*$ & $155.1$\\
$KD 3$  & $270$ & $260.01$ & $335$ & $328.6$ & $233$ & $244.1$ \\
$KD 4$  & $*$ & $428.2$ & $387$ & $383.6$ & $*$ & $398.1$ \\
$KD 5$  & $662$ & $668.8$ & $*$ & $832.6$ & $*$ & $562.1$ \\
$KD 6$  & $*$ & $701.3$ & $*$ & $923$& $*$ & $640.1$ \\
$KD 7$  & $*$ & $823.6$ & $*$ & $1100$& $*$ & $738.1$ \\
\hline
\end{tabular}
\flushleft{$^m$ Observed from INS or IRMS
\\$^c$ PCF Calculation}
\caption{Observed and calculated Eigen energies of the different materials studied. All units are in meV.}
\label{tab:fit1}
\end{table}

\begin{sidewaystable}[h!]
\centering
\caption{Eigenvectors and Eigenvalues for \textbf{0-Pr} in $|m_l,m_s\rangle$}
\setlength{\tabcolsep}{4pt}
\begin{tabular}{c|cccccccccccccc}
E (meV) &$|-3,-\frac{1}{2}\rangle$ & $|-3,\frac{1}{2}\rangle$ & $|-2,-\frac{1}{2}\rangle$ & $|-2,\frac{1}{2}\rangle$ & $|-1,-\frac{1}{2}\rangle$ & $|-1,\frac{1}{2}\rangle$ & $|0,-\frac{1}{2}\rangle$ & $|0,\frac{1}{2}\rangle$ & $|1,-\frac{1}{2}\rangle$ & $|1,\frac{1}{2}\rangle$ & $|2,-\frac{1}{2}\rangle$ & $|2,\frac{1}{2}\rangle$ & $|3,-\frac{1}{2}\rangle$ & $|3,\frac{1}{2}\rangle$ \tabularnewline
 \hline 
0.000 & 0.0 & 0.241 & -0.331 & 0.0 & 0.0 & 0.0 & 0.0 & 0.0 & 0.0 & -0.363 & 0.837 & 0.0 & 0.0 & 0.0 \tabularnewline
0.000 & 0.0 & 0.0 & 0.0 & -0.837 & 0.363 & 0.0 & 0.0 & 0.0 & 0.0 & 0.0 & 0.0 & 0.331 & -0.241 & 0.0 \tabularnewline
255.180 & 0.0 & 0.0 & 0.0 & 0.0 & 0.0 & 0.87 & -0.429 & 0.0 & 0.0 & 0.0 & 0.0 & 0.0 & 0.0 & -0.243 \tabularnewline
255.180 & 0.243 & 0.0 & 0.0 & 0.0 & 0.0 & 0.0 & 0.0 & 0.429 & -0.87 & 0.0 & 0.0 & 0.0 & 0.0 & 0.0 \tabularnewline
260.010 & 0.0 & 0.665 & -0.6 & 0.0 & 0.0 & 0.0 & 0.0 & 0.0 & 0.0 & -0.037 & -0.444 & 0.0 & 0.0 & 0.0 \tabularnewline
260.010 & 0.0 & 0.0 & 0.0 & -0.444 & -0.037 & 0.0 & 0.0 & 0.0 & 0.0 & 0.0 & 0.0 & -0.6 & 0.665 & 0.0 \tabularnewline
428.220 & 0.0 & 0.0 & 0.0 & -0.122 & -0.562 & 0.0 & 0.0 & 0.0 & 0.0 & 0.0 & 0.0 & 0.66 & 0.483 & 0.0 \tabularnewline
428.220 & 0.0 & -0.483 & -0.66 & 0.0 & 0.0 & 0.0 & 0.0 & 0.0 & 0.0 & 0.562 & 0.122 & 0.0 & 0.0 & 0.0 \tabularnewline
668.860 & 0.0 & 0.0 & 0.0 & -0.295 & -0.742 & 0.0 & 0.0 & 0.0 & 0.0 & 0.0 & 0.0 & -0.309 & -0.517 & 0.0 \tabularnewline
668.860 & 0.0 & -0.517 & -0.309 & 0.0 & 0.0 & 0.0 & 0.0 & 0.0 & 0.0 & -0.742 & -0.295 & 0.0 & 0.0 & 0.0 \tabularnewline
701.310 & -0.671 & 0.0 & 0.0 & 0.0 & 0.0 & 0.0 & 0.0 & 0.722 & 0.17 & 0.0 & 0.0 & 0.0 & 0.0 & 0.0 \tabularnewline
701.310 & 0.0 & 0.0 & 0.0 & 0.0 & 0.0 & 0.17 & 0.722 & 0.0 & 0.0 & 0.0 & 0.0 & 0.0 & 0.0 & -0.671 \tabularnewline
823.600 & 0.701 & 0.0 & 0.0 & 0.0 & 0.0 & 0.0 & 0.0 & 0.542 & 0.463 & 0.0 & 0.0 & 0.0 & 0.0 & 0.0 \tabularnewline
823.600 & 0.0 & 0.0 & 0.0 & 0.0 & 0.0 & -0.463 & -0.542 & 0.0 & 0.0 & 0.0 & 0.0 & 0.0 & 0.0 & -0.701 \tabularnewline
\end{tabular}
\label{flo:Eigenvectors}
\end{sidewaystable}

\clearpage

\begin{sidewaystable}[h!]
\centering
\caption{Eigenvectors and Eigenvalues for \textbf{1-Pr} in $|m_l,m_s\rangle$}
\setlength{\tabcolsep}{4pt}
\begin{tabular}{c|cccccccccccccc}
E (meV) &$|-3,-\frac{1}{2}\rangle$ & $|-3,\frac{1}{2}\rangle$ & $|-2,-\frac{1}{2}\rangle$ & $|-2,\frac{1}{2}\rangle$ & $|-1,-\frac{1}{2}\rangle$ & $|-1,\frac{1}{2}\rangle$ & $|0,-\frac{1}{2}\rangle$ & $|0,\frac{1}{2}\rangle$ & $|1,-\frac{1}{2}\rangle$ & $|1,\frac{1}{2}\rangle$ & $|2,-\frac{1}{2}\rangle$ & $|2,\frac{1}{2}\rangle$ & $|3,-\frac{1}{2}\rangle$ & $|3,\frac{1}{2}\rangle$ \tabularnewline
 \hline 
0.000 & 0.0 & 0.41 & -0.257 & 0.0 & 0.0 & 0.0 & 0.0 & 0.0 & 0.0 & -0.35 & 0.802 & 0.0 & 0.0 & 0.0 \tabularnewline
0.000 & 0.0 & 0.0 & 0.0 & -0.802 & 0.35 & 0.0 & 0.0 & 0.0 & 0.0 & 0.0 & 0.0 & 0.257 & -0.41 & 0.0 \tabularnewline
160.800 & 0.0 & -0.655 & 0.488 & 0.0 & 0.0 & 0.0 & 0.0 & 0.0 & 0.0 & 0.151 & 0.557 & 0.0 & 0.0 & 0.0 \tabularnewline
160.800 & 0.0 & 0.0 & 0.0 & -0.557 & -0.151 & 0.0 & 0.0 & 0.0 & 0.0 & 0.0 & 0.0 & -0.488 & 0.655 & 0.0 \tabularnewline
328.690 & -0.628 & 0.0 & 0.0 & 0.0 & 0.0 & 0.0 & 0.0 & -0.227 & 0.744 & 0.0 & 0.0 & 0.0 & 0.0 & 0.0 \tabularnewline
328.690 & 0.0 & 0.0 & 0.0 & 0.0 & 0.0 & 0.744 & -0.227 & 0.0 & 0.0 & 0.0 & 0.0 & 0.0 & 0.0 & -0.628 \tabularnewline
383.680 & 0.0 & -0.456 & -0.827 & 0.0 & 0.0 & 0.0 & 0.0 & 0.0 & 0.0 & 0.312 & 0.104 & 0.0 & 0.0 & 0.0 \tabularnewline
383.680 & 0.0 & 0.0 & 0.0 & -0.104 & -0.312 & 0.0 & 0.0 & 0.0 & 0.0 & 0.0 & 0.0 & 0.827 & 0.456 & 0.0 \tabularnewline
832.670 & -0.697 & 0.0 & 0.0 & 0.0 & 0.0 & 0.0 & 0.0 & 0.589 & -0.409 & 0.0 & 0.0 & 0.0 & 0.0 & 0.0 \tabularnewline
832.670 & 0.0 & 0.0 & 0.0 & 0.0 & 0.0 & -0.409 & 0.589 & 0.0 & 0.0 & 0.0 & 0.0 & 0.0 & 0.0 & -0.697 \tabularnewline
923.240 & 0.0 & 0.0 & 0.0 & -0.189 & -0.87 & 0.0 & 0.0 & 0.0 & 0.0 & 0.0 & 0.0 & -0.108 & -0.442 & 0.0 \tabularnewline
923.240 & 0.0 & 0.442 & 0.108 & 0.0 & 0.0 & 0.0 & 0.0 & 0.0 & 0.0 & 0.87 & 0.189 & 0.0 & 0.0 & 0.0 \tabularnewline
1100.120 & 0.0 & 0.0 & 0.0 & 0.0 & 0.0 & 0.528 & 0.776 & 0.0 & 0.0 & 0.0 & 0.0 & 0.0 & 0.0 & 0.346 \tabularnewline
1100.120 & 0.346 & 0.0 & 0.0 & 0.0 & 0.0 & 0.0 & 0.0 & 0.776 & 0.528 & 0.0 & 0.0 & 0.0 & 0.0 & 0.0 \tabularnewline
\end{tabular}
\label{flo:Eigenvectors1}
\end{sidewaystable}

\clearpage

\begin{sidewaystable}[h!]
\centering
\caption{Eigenvectors and Eigenvalues for \textbf{2-Pr} in $|m_l,m_s\rangle$}
\setlength{\tabcolsep}{4pt}
\begin{tabular}{c|cccccccccccccc}
E (meV) &$|-3,-\frac{1}{2}\rangle$ & $|-3,\frac{1}{2}\rangle$ & $|-2,-\frac{1}{2}\rangle$ & $|-2,\frac{1}{2}\rangle$ & $|-1,-\frac{1}{2}\rangle$ & $|-1,\frac{1}{2}\rangle$ & $|0,-\frac{1}{2}\rangle$ & $|0,\frac{1}{2}\rangle$ & $|1,-\frac{1}{2}\rangle$ & $|1,\frac{1}{2}\rangle$ & $|2,-\frac{1}{2}\rangle$ & $|2,\frac{1}{2}\rangle$ & $|3,-\frac{1}{2}\rangle$ & $|3,\frac{1}{2}\rangle$ \tabularnewline
 \hline 
0.000 & 0.0 & -0.418 & 0.325 & 0.0 & 0.0 & 0.0 & 0.0 & 0.0 & 0.0 & 0.363 & -0.767 & 0.0 & 0.0 & 0.0 \tabularnewline
0.000 & 0.0 & 0.0 & 0.0 & 0.767 & -0.363 & 0.0 & 0.0 & 0.0 & 0.0 & 0.0 & 0.0 & -0.325 & 0.418 & 0.0 \tabularnewline
155.190 & 0.0 & 0.749 & -0.357 & 0.0 & 0.0 & 0.0 & 0.0 & 0.0 & 0.0 & 0.004 & -0.557 & 0.0 & 0.0 & 0.0 \tabularnewline
155.190 & 0.0 & 0.0 & 0.0 & 0.557 & -0.004 & 0.0 & 0.0 & 0.0 & 0.0 & 0.0 & 0.0 & 0.357 & -0.749 & 0.0 \tabularnewline
244.090 & 0.338 & 0.0 & 0.0 & 0.0 & 0.0 & 0.0 & 0.0 & 0.475 & -0.813 & 0.0 & 0.0 & 0.0 & 0.0 & 0.0 \tabularnewline
244.090 & 0.0 & 0.0 & 0.0 & 0.0 & 0.0 & 0.813 & -0.475 & 0.0 & 0.0 & 0.0 & 0.0 & 0.0 & 0.0 & -0.338 \tabularnewline
398.380 & 0.0 & 0.0 & 0.0 & -0.025 & -0.365 & 0.0 & 0.0 & 0.0 & 0.0 & 0.0 & 0.0 & 0.846 & 0.387 & 0.0 \tabularnewline
398.380 & 0.0 & -0.387 & -0.846 & 0.0 & 0.0 & 0.0 & 0.0 & 0.0 & 0.0 & 0.365 & 0.025 & 0.0 & 0.0 & 0.0 \tabularnewline
562.960 & 0.0 & 0.0 & 0.0 & 0.0 & 0.0 & 0.039 & -0.534 & 0.0 & 0.0 & 0.0 & 0.0 & 0.0 & 0.0 & 0.845 \tabularnewline
562.960 & -0.845 & 0.0 & 0.0 & 0.0 & 0.0 & 0.0 & 0.0 & 0.534 & -0.039 & 0.0 & 0.0 & 0.0 & 0.0 & 0.0 \tabularnewline
640.850 & 0.0 & 0.338 & 0.224 & 0.0 & 0.0 & 0.0 & 0.0 & 0.0 & 0.0 & 0.857 & 0.317 & 0.0 & 0.0 & 0.0 \tabularnewline
640.850 & 0.0 & 0.0 & 0.0 & -0.317 & -0.857 & 0.0 & 0.0 & 0.0 & 0.0 & 0.0 & 0.0 & -0.224 & -0.338 & 0.0 \tabularnewline
738.740 & 0.0 & 0.0 & 0.0 & 0.0 & 0.0 & -0.581 & -0.7 & 0.0 & 0.0 & 0.0 & 0.0 & 0.0 & 0.0 & -0.415 \tabularnewline
738.740 & -0.415 & 0.0 & 0.0 & 0.0 & 0.0 & 0.0 & 0.0 & -0.7 & -0.581 & 0.0 & 0.0 & 0.0 & 0.0 & 0.0 \tabularnewline
\end{tabular}
\label{flo:Eigenvectors2}
\end{sidewaystable}

\clearpage
\section{First-principles calculations.}
In a first step, sets of scalar-relativistic (SR) multiconfigurational wavefunctions were calculated with the complete active space (CAS) self-consistent field approach\cite{roos1980b, olsen1988, Malmqvist:1990a}. Subsequently, these wavefunctions were employed in single-state, multireference CAS second-order perturbation theory (PT2)\cite{andersson1990, angeli_introduction_2001} calculations in order to obtain more accurate energies including effects from dynamic correlation. An imaginary shift of 0.20 au was used with PT2 in order to minimize intruder-state effects. SR effects were introduced with the second-order Douglas-Kroll-Hess Hamiltonian.\cite{Douglas:1974a, Hess:1985a, Hess:1986, Wolf:2002a} In a second step, spin-orbit (SO) coupling was introduced using the restricted active-space state-interaction (RASSI) formalism\cite{malmqvist02cpl} and the atomic mean-field integrals (AMFI).\cite{Hess:1996a} Henceforth, PT2-SR and PT2-SO labels will be used to identify results obtained with only a SR treatment (`spin-free' states belonging to a well-defined spin multiplicity) or with both SR and SO coupling treatments.

\begin{figure}[h!]
    \begin{center}
        \includegraphics[width=1.\textwidth]{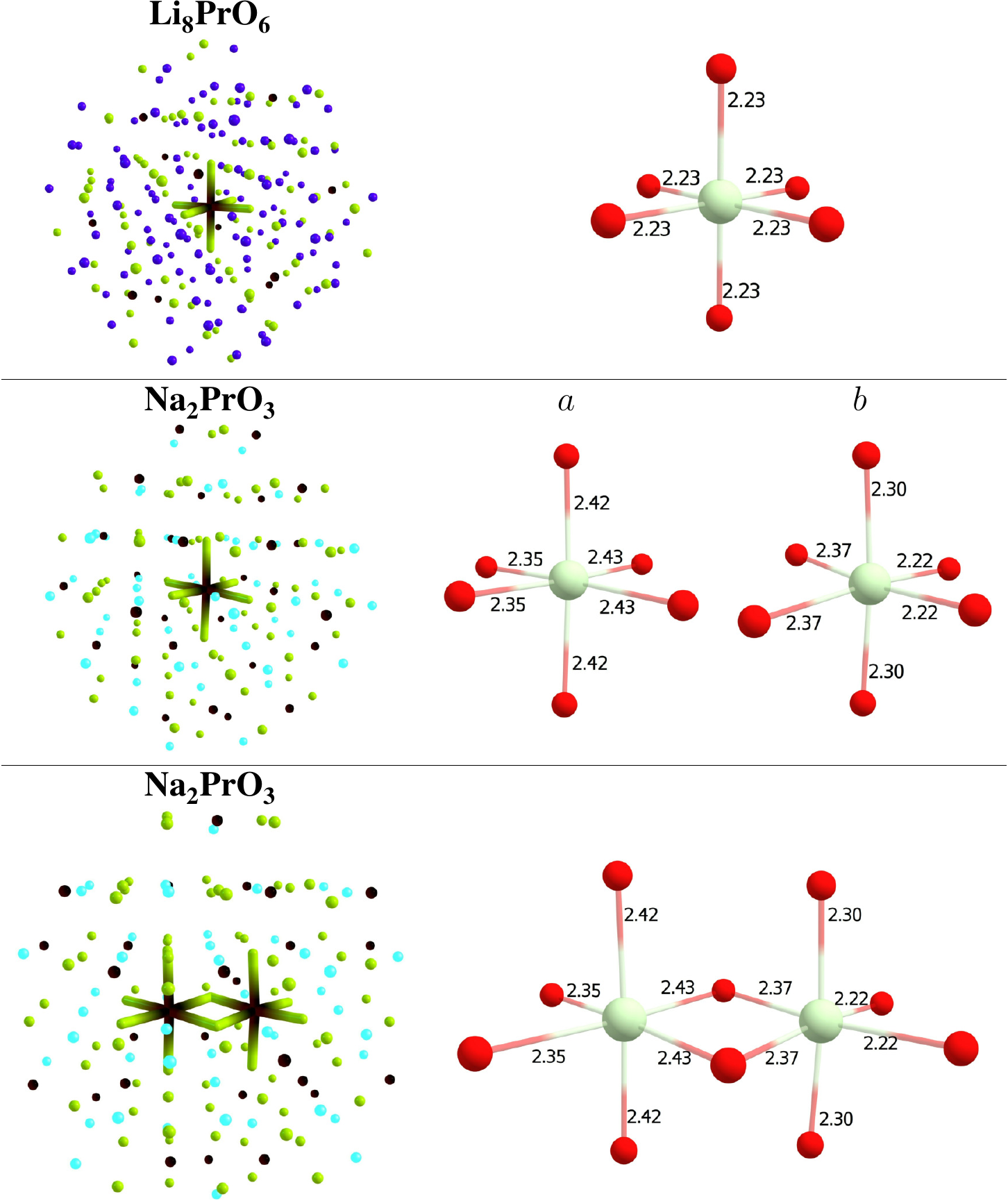}
    \end{center}
    \caption{\label{fig:geoms}
{\bf Embedded cluster models and bare ion geometries used in the present study:} {[PrO$_6$]$^{8-}$} and {[Pr$_2$O$_{10}$]$^{12-}$} with surrounding embedding pseudocharges (left panels) and without (right panels).
 	}    
\end{figure}

Embedding was achieved using a similar strategy as in our previous work on actinide and lanthanide systems.\cite{Autschbach:2016v, Autschbach:2021a} Geometries for the systems under investigation are shown in Fig S\ref{fig:geoms}. Within \textbf{0-Pr}, the [PrO$_6$]$^{8-}$ ion is isolated by Li$^+$ cations and adopts a nearly octahedral geometry ($O_h$) with a Pr--O bond length of 2.32 \AA{} and bond angles within 90$\pm$2.5$^{\circ}$. In \textbf{2-Pr}, edge-shared {[Pr$_2$O$_10$]$^{12-}$} dimers occur, in a honeycomb lattice, with structures obeying $C_2$ symmetry. The individual {[PrO$_6$]$^{12-}$} monomers, labeled with $a$ and $b$ in center panel of Fig S\ref{fig:geoms}, both comply with $C_2$ local symmetry and exhibit geometries that are significantly different between themselves and significantly distorted from $O_h$. For instance, the Pr--O bond lengths are more than 0.1 \AA{} larger in monomer $a$ vs.\ $b$, and the bond angles vary around 90$\pm$10$^\circ$ in both monomers.    
    
\begin{figure}[h!]
    \begin{center}
        \includegraphics[width=.9\textwidth]{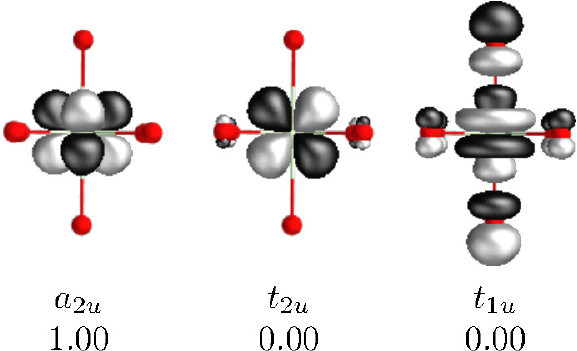}
    \end{center}
    \caption{\label{fig:ohorbs}
{\bf Ground-state natural orbitals:} Ground-state natural orbitals ($\pm$0.03 au isosurfaces) of Pr $4f$ AO parentage and corresponding populations calculated with CAS-SR for an $O_h$ {[PrO$_6$]$^{8-}$} isolated ion with a Pr--O bond length of 2.23 \AA{}.).
 	}    
\end{figure}
    
\begin{table}[h!]
\centering
\caption{Electronic structure of the {Pr$^{4+}$} ion.}
\renewcommand{\arraystretch}{2}
\begin{tabular}{ccccc}
\hline
System & PT2-SR & $\Delta E$ & PT2-SO & $\Delta E$ \\ \hline

{Pr$^{4+}$} & $^2F$ & 0.000 & $^2F_{5/2}$ & 0.000  \\  
             &          &       & $^2F_{7/2}$ & 395  \\ 
\hline
\end{tabular}
\label{tab:pr}
\end{table}

\begin{figure}[h!]
\centering
\includegraphics[width=.8\textwidth]{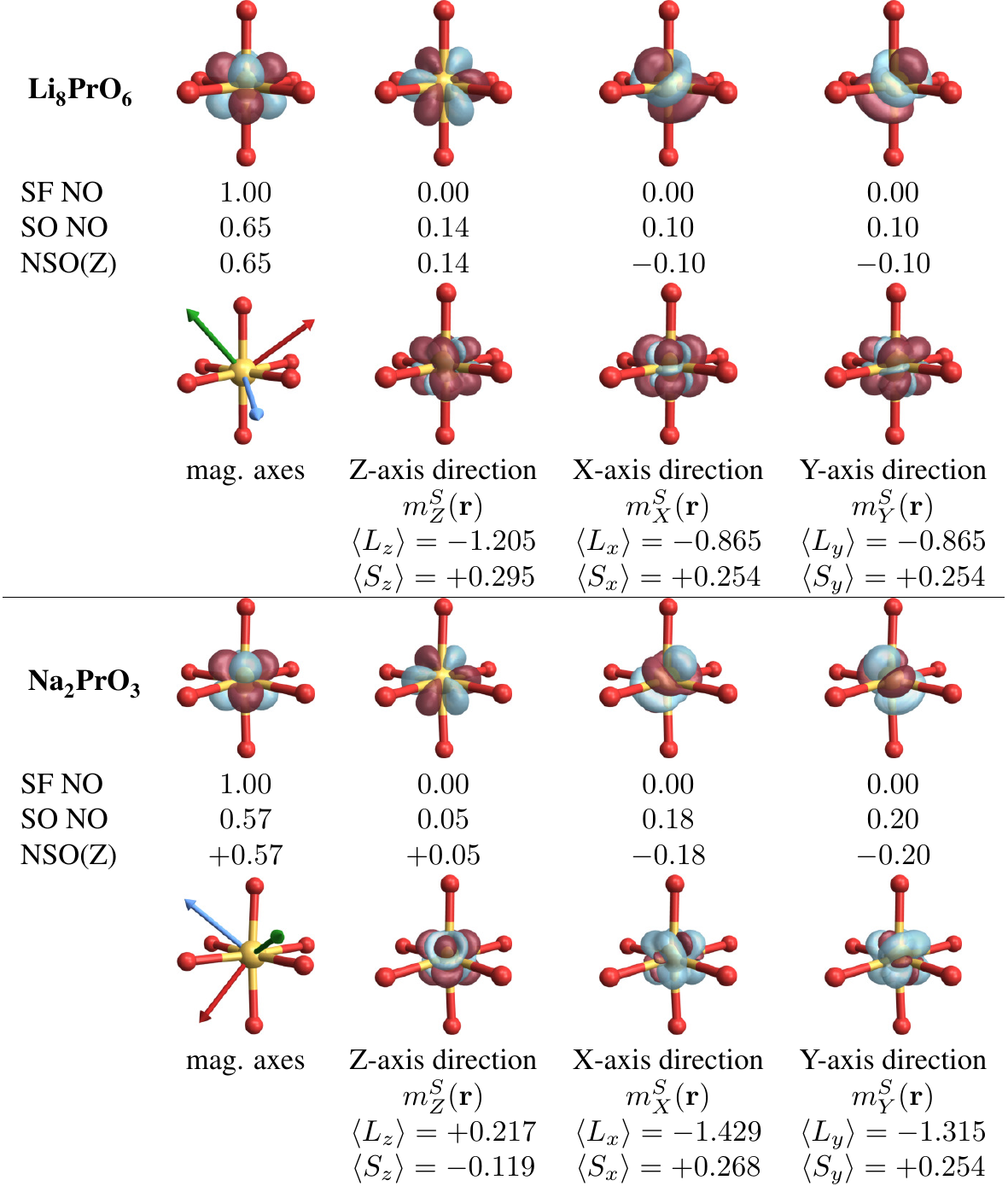}
\caption{Selected natural orbitals (NOs, $\pm0.03$ au isosurfaces) and their populations, with and without SO coupling, calculated for the {[PrO$_6$]$^{8-}$} embedded cluster model of \textbf{0-Pr} (top panel) and \textbf{2-Pr} (bottom panel). Also given are the spin populations corresponding to the natural spin orbitals (NSOs, isosurfaces identical with those of the NOs) calculated for the Z direction of the spin magnetization, along the magnetic Z-axis, for the GS Kramers component with $\langle S_{z} \rangle > 0$. Plots of the spin magnetization [$m_W^S(\mathbf{r})$] for quantization along the $W = X$, $Y$ and $Z$ magnetic axes are also given with isosurface of $\pm$0.001 au. Color code for magnetic axes: red for Z-axis, green for X-axis, blue for Y-axis.}
\label{fig:pro-orbs}
\end{figure}

Calculated data for the low-energy electronic states of {[PrO$_6$]$^{12-}$}, with PT2-SR and PT2-SO, are collected in Table S\ref{tab:pr-pro6oh} (isolated $O_h$ structure), Table S\ref{tab:pro6-li} (structure from 0-Pr) and Table S\ref{tab:pro6-na} (structures from 2-Pr). Selected natural orbital (NO), natural spin-magnetization orbital (NSO) and spin magnetization plots [$m_W^S(\mathbf{r})$ for quantization along the $W = X$, $Y$ and $Z$ magnetic axes] are shown in Fig S\ref{fig:pro-orbs}. Details regarding the generation of NSOs and $m_W^S(\mathbf{r})$ can be found in Reference \cite{Autschbach:2014a}. Note that the NSO spin populations add up to twice the spin expectation value for a given quantization axis, $2\langle S_W \rangle$, while $m_W^S(\mathbf{r})$ corresponds to the usual spin density if it is calculated without SO coupling.  Regardless of the {[PrO$_6$]$^{12-}$} structure considered, the spin-free GS is $^2A_{2u}$ (or of $^2A_{2u}$ parentage for the non-$O_h$ cases) with a single configurational wavefunction derived from the 4f $a_{2u}^1$ configuration.     

\begin{figure}[h!]
\centering
\includegraphics[width=\textwidth]{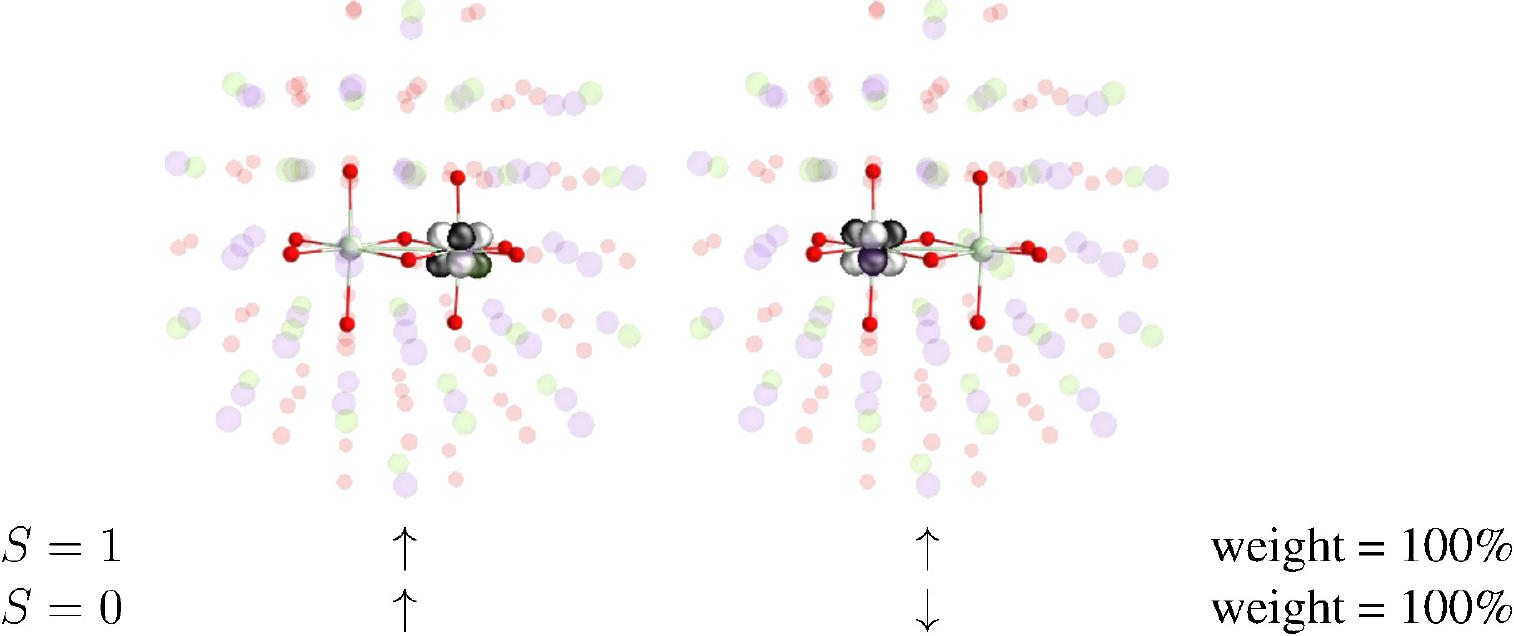}
\caption{Localized CAS orbitals characterizing the electron pairing in the lowest-energy spin-triplet ($S = 1$) and spin-singlet ($S = 0$) configurations of {[Pr$_2$O$_{10}$]$^{12-}$} embedded-cluster model of {Na$_2$PrO$_3$}. The localization was achieved by arbitrary rotations among the CAS NOs and a subsequent CAS configuration interaction was performed to tailor the wavefunction configurational admixture.}
\label{fig:binuc-sforbs}
\end{figure}

\begin{figure}[h!]
\centering
\includegraphics[width=\textwidth]{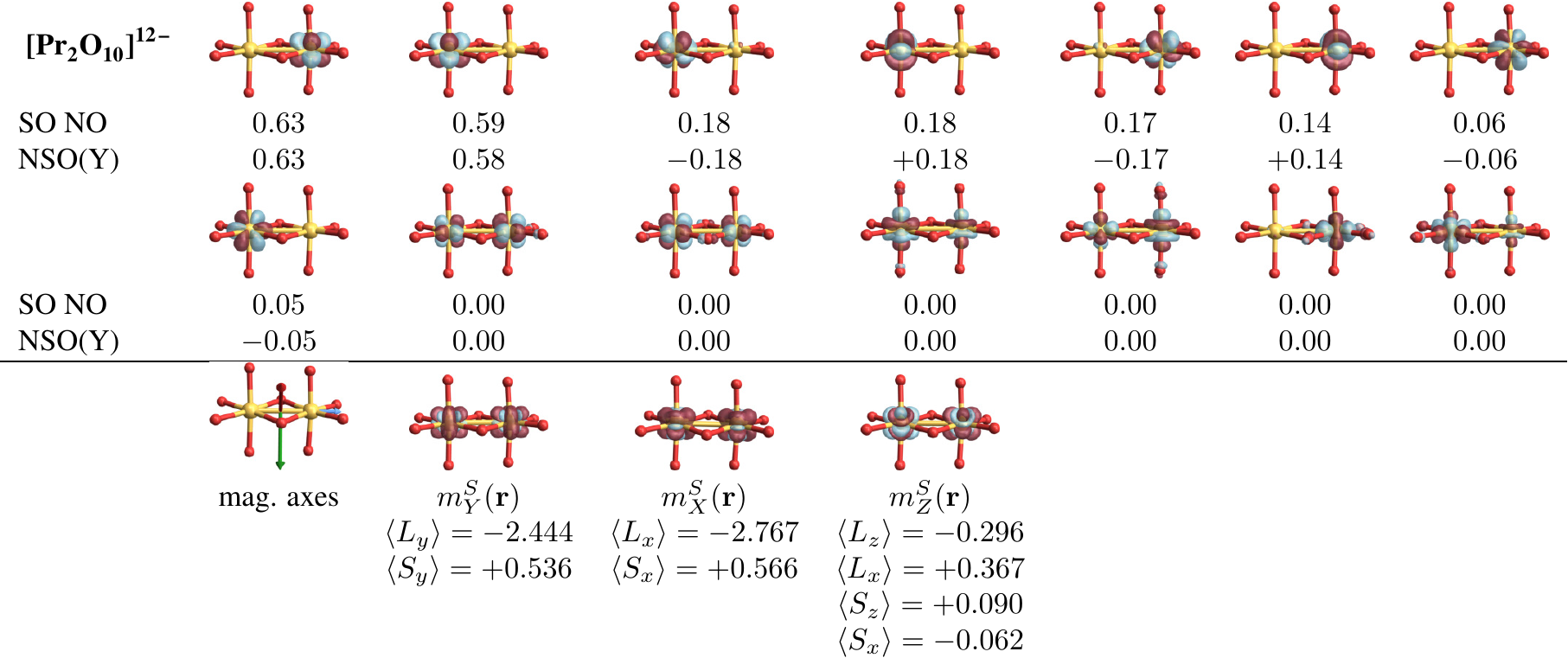}
\caption{Top panel: Selected natural orbitals (NOs, $\pm0.03$ au isosurfaces) and their populations calculated with PT2-SO for the {[Pr$_2$O$_{10}$]$^{12-}$} embedded cluster model of \textbf{2-Pr}. Also given are the spin populations corresponding to the natural spin orbitals (NSOs, isosurfaces identical with those of the NOs) calculated for the Y direction of the spin magnetization, along the magnetic Y-axis. Bottom panel: Plots of the spin magnetization [$m_W^S(\mathbf{r})$] for quantization along the $W = X$, $Y$ and $Z$ magnetic axes are also given with isosurface of $\pm$0.001 au. Color code for magnetic axes: red for Z-axis, green for X-axis, blue for Y-axis.}
\label{fig:binuc-orbs}
\end{figure}

Data obtained for the binuclear {[Pr$_2$O$_{10}$]$^{12-}$} embedded-cluster model of 2-Pr are gathered in Table S\ref{tab:pro6-binuc} (energies and magnetism), Fig S\ref{fig:binuc-sforbs} (localized CAS NOs showing the magnetic orbitals), and in Fig S\ref{fig:binuc-orbs} (PT2-SO NOs and populations, NSOs and spin magnetization plots). The lowest energy $S = 1$ spin-triplet and $S = 0$ spin-singlet states are generated by the spin pairing of the 4f $a_{2u}^1$ electrons localized at the two Pr$^{4+}$ centers (Fig S\ref{fig:binuc-sforbs}). The resulting wavefunctions are single-configurational and described by the Pr$_1$ $a_{2u}^\uparrow$ + Pr$_2$ $a_{2u}^\uparrow$ ($S = 1$) and Pr$_1$ $a_{2u}^\uparrow$ + Pr$_2$ $a_{2u}^\downarrow$ ($S = 0$) configurations. The energy difference obtained with PT2-SR, which identifies with the Heisenberg exchange coupling ($J$), is 4.2 meV. With SO coupling, there are four low-energy states split by 1.5 meV. These states can be regarded to originate from the GS Kramers doublets of the individual mononuclear systems. Analysis of the SO wavefunctions showed that the lowest energy SO state has the largest (one-state) contribution (37\%) from the lowest energy $S = 0$ spin-free state while the next three SO states with energies ranging between 0.7--1.5 meV have largest (one-state) contributions (also 37\%) from the lowest-energy $S = 1$ spin-free state. The strong SO coupling in the GS of the complex is reflected in the NO and NSO populations shown in Fig S\ref{fig:binuc-orbs}, and also in the small spin expectation value ($\langle S_y \rangle = 0.536$) along the Pr--Pr internuclear axis (magnetic Y-axis, the magnetic axes are also depicted in Fig S\ref{fig:binuc-orbs}).

 
\begin{table}[h!]
\centering
\caption{Electronic structure of an $O_h$ {[PrO$_6$]$^{8-}$} isolated free ion.}
\begin{tabular}{ccccccccc}
\hline
Atomic     & \multicolumn{3}{c}{Spin-free states} & $J$  & \multicolumn{3}{c}{Kramers doublets} \\ 
multiplet &  & $\Delta E^b$ & $\Delta E^c$ & multiplet & SF comp.  & $\Delta E^b$ & $\Delta E^c$  \\ \hline

$^2F$                  & $^2A_{2u}$ & 0.000 & 0.000 & $^2F_{5/2}$ & 58$^2A_{2u}$ + 42$^2T_{2u}$ & 0.000 & 0.000 \\
                       & $^2T_{2u}$ & 115 & 170 & & 54$^2T_{2u}$ + 44$^2T_{1u}$ & 135 & 143 \\
                       & $^2T_{1u}$ & 252 & 256 & &  54$^2T_{2u}$ + 44$^2T_{1u}$ & 135 & 143 \\ \\       
                       &            &       & & $^2F_{7/2}$ & 58$^2T_{2u}$ + 42$^2A_{2u}$  & 385 & 397 \\
                       &            &       & & & 53$^2T_{1u}$ + 46$^2T_{2u}$ & 503 & 511 \\
                       &            &       & & & 50$^2T_{2u}$ + 46$^2T_{1u}$ & 503 & 511 \\                     
                       &            &       & & & 100$^2T_{2u}$ & 576 & 558 \\

\hline

\multicolumn{8}{c}{Main values of the $g$ tensor in the ground Kramers doublet} \\
$g_X$ & & & & & & 1.116 & 0.906 \\
$g_Y$ & & & & & & 1.116 & 0.906 \\
$g_Z$ & & & & & & 1.116 & 0.906 \\ \hline
\end{tabular}
\flushleft{$^a$The $O_h$ geometry was obtained by symmetrizing the {PrO$_6$} unit of {Li$_8$PrO$_6$}. All energies are in meV units.
\\$^b$State-averaged calculations without symmetry constraint.
\\$^c$State-averaged calculations with $D_{2h}$ symmetry constraint.}
\label{tab:pr-pro6oh}
\end{table}



\begin{table}[h!]
\centering
\caption{Electronic structure of a {[PrO$_6$]$^{8-}$} unit of \textbf{0-Pr}.$^a$}
\begin{tabular}{cccccccccc}
\hline
Atomic & \multicolumn{3}{c}{Spin-free states} & $J$  & \multicolumn{3}{c}{Kramers doublets} \\ 
multiplet & & $\Delta E^b$ & $\Delta E^c$ & multiplet & SF comp.$^c$ & $\Delta E^b$ & $\Delta E^c$  \\ \hline

$^2F$ & $^2A_{2u}$ & 0.000 & 0.000 & $^2F_{5/2}$ & 65$^2A_{2u}$ + 35$^2T_{2u}$ & 0.000 & 0.000 \\ 
      & $^2T_{2u}$ & 129 & 158 & & 76$^2T_{2u}$ + 24$^2T_{1u}$ & 141 & 241 \\
      &           & 129 & 184 & & 80$^2T_{2u}$ + 20$^2T_{1u}$ & 151 & 246 \\
      
      &          & 148 & 184 & \\ \\
 
 & $^2T_{1u}$ & 270 & 498 & $^2F_{7/2}$ & 65$^2T_{2u}$ + 35$^2A_{2u}$  & 388 & 396 \\ 
 &            & 272 & 544 & & 65$^2T_{1u}$ + 20$^2T_{2u}$ & 513 & 662 \\ 
 &            &  273 & 544 & & 80$^2T_{1u}$ + 20$^2T_{2u}$ & 516 & 707 \\                                                                   
  & &           &       & & 100$^2T_{1u}$ & 588 & 820 \\                                                                    
\hline

\multicolumn{8}{c}{Main values of the $g$ tensor in the ground Kramers doublet} \\
$g_X$ & & & & & & 1.015 & 0.714  \\
$g_Y$ & & & & & & 1.018 & 0.714  \\
$g_Z$ & & & & & & 1.095 & 1.229  \\ \hline

\end{tabular}

\flushleft{$^a$All energies are in eV units. Selected orbital isosurfaces and populations are shown in Figure \ref{fig:pro-orbs}. \\
$^b${[PrO$_6$]$^{8-}$} isolated ion of Fig S\ref{fig:geoms} (top panel). \\
$^c${[PrO$_6$]$^{8-}$} embedded-cluster model of Fig S\ref{fig:geoms} (top panel). \\
$^d$Magnetic axes are depicted in Fig S\ref{fig:pro-orbs}.}
\label{tab:pro6-li}
\end{table}



\begin{table}[htbp]
\centering
\caption{Electronic structure of the {[PrO$_6$]$^{8-}$} embedded crystal-model system of \textbf{2-Pr}.$^a$}
\small
\begin{tabular}{ccccccccccccc}
\hline
Atomic & \multicolumn{4}{c}{Spin-free states} & $J$  & \multicolumn{4}{c}{Kramers doublets} \\ 
multiplet &   & $\Delta E^b$ & $\Delta E^c$ & $\Delta E^d$ & multiplet & SF comp.$^d$ &  $\Delta E^b$ & $\Delta E^c$ & $\Delta E^d$  \\ \hline

$^2F$ & $^2A_{2u}$ & 0.000 & 0.000 & 0.000  & $^2F_{5/2}$ & 57$^2A_{2u}$ + 43$^2T_{2u}$ & 0.000 & 0.000 & 0.000 \\ 
      & & & & & & 63$^2T_{2u}$ + 37$^2T_{1u}$ & 118 & 136 & 168 \\
      & $^2T_{2u}$ & 089 & 118 & 096  & & 69$^2T_{2u}$ + 31$^2T_{1u}$ & 162 & 163 & 231 \\
      &            & 099 & 120 & 099  & & & & &  \\
      &            & 156 & 159 & 201  & $^2F_{7/2}$ & 59$^2T_{2u}$ + 38$^2A_{2u}$ & 387 & 389 & 394  \\
      & & & & & & 63$^2T_{1u}$ + 37$^2T_{2u}$ & 479 & 498 & 536 \\
      & $^2T_{1u}$ & 214 & 248 & 323 & & 77$^2T_{1u}$ + 23$^2T_{2u}$ & 529 & 537  & 623   \\ 
      &            & 237 & 268 & 349 & & 94$^2T_{1u}$ + 6$^2T_{2u}$ & 597 & 601 & 754 \\ 
      &            & 306 & 312 & 502 \\                                                                   

\hline
\multicolumn{10}{c}{Main values of the $g$ tensor in the ground Kramers doublet$^d$} \\
$g_X$ & & & & & & & 1.694 & 1.490 & 1.772  \\
$g_Y$ & & & & & & & 1.649 & 1.310 & 1.536  \\
$g_Z$ & & & & & & & 0.110 & 0.406 & 0.049  \\

\multicolumn{10}{c}{Expectation values$^d$} \\
\multicolumn{10}{c}{Magnetic $X$-axis direction$^e$} \\
$\langle L_{x} \rangle$ & & & & & & & $-$1.359 & $-$1.265 & $-$1.429 \\
$\langle S_{x} \rangle$ & & & & & & & 0.256 & 0.260 & 0.268 \\ 

\multicolumn{10}{c}{Magnetic $Y$-axis direction$^e$} \\
$\langle L_{y} \rangle$ & & & & & & & $-$1.312 & $-$1.155 & $-$1.315 \\
$\langle S_{y} \rangle$ & & & & & & & 0.243 & 0.250 & 0.254 \\ 

\multicolumn{10}{c}{Magnetic $Z$-axis direction$^e$} \\
$\langle L_{z} \rangle$ & & & & & & & $-$0.306 & $-$0.558 & 0.217 \\
$\langle S_{z} \rangle$ & & & & & & & 0.125 & 0.177 & $-$0.119 \\ \hline

\end{tabular}

\flushleft{$^a$All energies are in eV units. Selected orbital isosurfaces and populations are shown in Figure S\ref{fig:pro-orbs}. \\ 
$^b$Structure $a$ in Fig S\ref{fig:geoms}. \\
$^c$Structure $b$ in Fig S\ref{fig:geoms}.\\
$^d$Structure $a$ plus embedding in Fig S\ref{fig:geoms}. \\
$^e$Magnetic axes are depicted in Fig S\ref{fig:pro-orbs}.}
\label{tab:pro6-na}
\end{table}

\begin{table}[ht!]
\centering
\caption{Electronic structure of the [Pr$_2$O$_{10}$]$^{12-}$ embedded crystal-model system of \textbf{2-Pr}.}
\begin{tabular}{cccccccccc}
\hline
Atomic & \multicolumn{2}{c}{Spin-free states} & $J$  & \multicolumn{2}{c}{Spin-orbit states} \\ 
multiplet &   & $\Delta E$ (eV) & multiplet &   & $\Delta E$ (eV)  \\ \hline

$^2F^{\text{Pr}_1}$ + $^2F^{\text{Pr}_2}$ & $S=0$ & 0.0000 & $^2F_{5/2}^{\text{Pr}_1}$ + $^2F_{5/2}^{\text{Pr}_2}$ & 37\% ($S=0$)  & 0.0000 \\ 
      & $S=1$ & 0.0042 &  & 37\% ($S=1$)  & 0.0007\\
      &               &  & & 37\% ($S=1$) & 0.0008 \\
      &               &  & & 37\% ($S=1$) & 0.0015 \\ \hline
\multicolumn{2}{c}{Heisenberg exchange coupling ($J$)} & 4.2 meV &  & & $\approx$1 meV$^a$\\ \hline

Expectation values \\ 
\multicolumn{2}{c}{Magnetic $X$-axis direction$^b$} \\
$\langle L_{x} \rangle$ & & &  & & $-$2.767 \\
$\langle S_{x} \rangle$ & & &  & & 0.566 \\

\multicolumn{2}{c}{Magnetic $Y$-axis direction$^b$} \\
$\langle L_{y} \rangle$ & & &  & & $-$2.444 \\
$\langle S_{y} \rangle$ & & &  & & 0.536 \\

\multicolumn{2}{c}{Magnetic $Z$-axis direction$^b$} \\
$\langle L_{z} \rangle$ & & &  & & $-$0.296 \\
$\langle L_{x} \rangle$ & & &  & & 0.367 \\
$\langle S_{z} \rangle$ & & &  & & 0.090 \\

\hline
\end{tabular}
\label{tab:pro6-binuc}
\flushleft{$^a$With SO coupling, spin is not a good quantum number. \\
$^b$Magnetic axes are depicted in Figure S\ref{fig:binuc-orbs}. The magnetic Y-axis corresponds to the Pr--Pr internuclear axis.}
\end{table}


\clearpage
\section{References}
\renewcommand\refname{\vskip -1cm}

\clearpage